\documentclass{sigplanconf}

\usepackage{amsmath, amsthm, amssymb}
\usepackage{graphicx}
\usepackage{url}
\usepackage{stmaryrd}
\usepackage{bussproofs}

\newcommand{\ignore}[1]{}
\newcommand{\ital}[1]{{\em #1}}

\newcommand{\eg}{{\em e.g.},}
\newcommand{\ra}{\rightarrow}

\newcommand{\simerr}{overflow}
\newcommand{\lprolog}{$\lambda$Prolog}
\newcommand{\hhf}{$hohh$}
\newcommand{\lftype}{\text{\ital{lf-type}}}
\newcommand{\lfobject}{\text{\ital{lf-obj}}}

\newcommand{\ruvappt}{{\scriptsize APP$_\text{t}$}}
\newcommand{\ruvpit}{{\scriptsize PI$_\text{t}$}}
\newcommand{\ruvinito}{{\scriptsize INIT$_\text{o}$}}
\newcommand{\ruvappo}{{\scriptsize APP$_\text{o}$}}
\newcommand{\ruvabso}{{\scriptsize ABS$_\text{o}$}}

\renewcommand{\vec}[1]{\overrightarrow{#1}}

\newcommand{\subst}[1]{[#1]}

\newcommand{\iprove}[2]{\sequent{#1}{#2}}
\newcommand{\lfprove}[2]{#1 \, \vdash \, #2}
\newcommand{\hastype}[2]{hastype \ #1 \ #2}

\newcommand{\encEq}[2]{#1 \sim #2}

\newcommand{\encTerm}[1]{\langle #1 \rangle}
\newcommand{\enc}[1]{\lbrace\!\!\lbrace #1 \rbrace\!\!\rbrace}
\newcommand{\encExtP}[2]{\llbracket #1 \rrbracket^{+}_{#2}}
\newcommand{\encExtN}[1]{\llbracket #1 \rrbracket^{-}}
\newcommand{\termUV}[2]{#1 \sqsubset_o #2}
\newcommand{\formulaUV}[2]{#1 \sqsubset_t #2}

\newcommand{\forallx}[2]{\forall #1.#2}
\newcommand{\sequent}[2]{#1 \longrightarrow #2}
\newcommand{\sigsequent}[3]{#1 ; #2 \longrightarrow #3}
\newcommand{\focusedsigseq}[4]{#1 ; #2 \buildrel #4 \over
  \longrightarrow #3}
\newcommand{\lambdax}[2]{\lambda #1.#2}
\newcommand{\typedlambda}[3]{\lambda #1 \mbox{:} #2 . #3}
\newcommand{\typedlambdas}[3]{\lambda \vec{#1 \mbox{:} #2} . #3}
\newcommand{\typedpi}[3]{\Pi #1 \mbox{:} #2 . #3}
\newcommand{\typedpis}[3]{\Pi \vec{#1 \mbox{:} #2} . #3}
\newcommand{\app}{\ }
\newcommand{\emptyctx}{\cdot}
\newcommand{\oftype}[2]{#1 : #2}
\newcommand{\ctx}{\mbox{\sl ctx}}
\newcommand{\kind}{\mbox{\sl kind}}
\newcommand{\type}{\mbox{\sl Type}}

\newcommand{\bcGoal}{\mbox{\sl backchain}}
\newcommand{\topGoal}{$\top \mbox{\sl R}$}

\newcommand{\impGoal}{$\supset\! \mbox{\sl R}$}
\newcommand{\allGoal}{$\forall \mbox{\sl R}$}
\newcommand{\atomicGoal}{$\mbox{\sl decide}$}
\newcommand{\initial}{$\mbox{\sl init}$}
\newcommand{\allClause}{$\forall \mbox{\sl L}$}
\newcommand{\impClause}{$\supset\! \mbox{\sl L}$}

\newcommand{\nullctx}{\mbox{\sl null-ctx}}
\newcommand{\kindctx}{\mbox{\sl kind-ctx}}
\newcommand{\typectx}{\mbox{\sl type-ctx}}
\newcommand{\typekind}{\mbox{\sl type-kind}}
\newcommand{\pikind}{\mbox{\sl pi-kind}}
\newcommand{\varfam}{\mbox{\sl var-fam}}
\newcommand{\varobj}{\mbox{\sl var-obj}}
\newcommand{\pifam}{\mbox{\sl pi-fam}}
\newcommand{\absfam}{\mbox{\sl abs-fam}}
\newcommand{\appfam}{\mbox{\sl app-fam}}
\newcommand{\appobj}{\mbox{\sl app-obj}}
\newcommand{\absobj}{\mbox{\sl abs-obj}}

\newcommand{\ie}{{\it i.e.}}

\newtheorem{definition}{Definition}

\newtheorem{theorem'}{Theorem}
\newtheorem{corollary'}{Corollary}
\newtheorem{lemma'}{Lemma}
\newtheorem{proposition}{Proposition}

\newenvironment{packed_itemize}{
\begin{itemize}
}{\end{itemize}}

\begin{document}

\conferenceinfo{PPDP'10,} {July 26--28, 2010, Hagenberg, Austria.}
\CopyrightYear{2010}
\crdata{978-1-4503-0132-9/10/07}

\title{A Meta-Programming Approach to Realizing\\
Dependently Typed Logic Programming}

\authorinfo{Zachary Snow}
           {Computer Science and Engineering\\
            University of Minnesota\\
            200 Union Street SE\\
            Minneapolis, MN 55455}
           {snow@cs.umn.edu}

\authorinfo{David Baelde}
           {Computer Science and Engineering\\
            University of Minnesota\\
            200 Union Street SE\\
            Minneapolis, MN 55455}
           {dbaelde@cs.umn.edu}

\authorinfo{Gopalan Nadathur}
           {Computer Science and Engineering\\
            University of Minnesota\\
            200 Union Street SE\\
            Minneapolis, MN 55455}
           {gopalan@cs.umn.edu}
 
\maketitle

\begin{abstract}
Dependently typed $\lambda$-calculi such as the Logical Framework (LF)
can encode relationships between terms in types and can naturally
capture correspondences between formulas and their proofs. Such
calculi can also be given a logic programming interpretation: the
Twelf system is based on such an interpretation of LF. We consider
here whether a conventional logic programming language can 
provide the benefits of a Twelf-like system
for encoding type and proof-and-formula dependencies.
In particular, we present a simple mapping from LF specifications to
a set of formulas in the higher-order hereditary Harrop (\hhf) language,
that relates derivations and proof-search between the two frameworks.
We then show that this encoding can be
improved by exploiting knowledge of the well-formedness of the
original LF specifications to elide much redundant type-checking
information. The resulting logic program has a structure that closely
resembles the original specification, thereby allowing LF specifications
to be viewed as \hhf\ meta-programs. Using the Teyjus implementation of
$\lambda$Prolog, we show that our translation provides an efficient
means for executing LF specifications, complementing the ability that
the Twelf system provides for reasoning about them.

\end{abstract}




\section{Introduction}
\label{sec:introduction}

There is a significant, and growing interest in mechanisms for
specifying, prototyping and reasoning about formal systems that are
described by syntax-directed rules. 
Dependently typed $\lambda$-calculi such as the Logical
Framework (LF) \cite{harper93jacm} provide many conveniences from a
specification perspective in this context. Such calculi facilitate the
use of a higher-order approach to describing the syntax 
of formal objects and they allow relationships between terms to be captured
in an elegant way through type dependencies.
Furthermore, dependently typed $\lambda$-calculi enjoy a well-known
isomorphism between formulas and types \cite{howard80}, leading to a
unification of the concept of a proof of a formula with an inhabitant of a given
type.  
Thus, the search for type inhabitants can be identified with
proof-search and can thereby be given a logic programming
interpretation. The Twelf system \cite{pfenning99cade} that we
consider here exploits these possibilities relative to LF. As
such, it has been used  
successfully in specifying and prototyping varied formal systems, and
mechanisms have also been built into it to reason about
specifications.

\sloppy

Predicate logics are also capable of encoding syntax-directed specifications,
and provide the basis for logic programming languages
in the familiar tradition of Prolog. Within this 
framework, the logic of higher-order hereditary Harrop (\hhf) formulas 
\cite{miller91apal} that underlies the language $\lambda$Prolog
\cite{nadathur88iclp} provides a builtin ability to treat
binding notions in syntax and thus has
particular usefulness in representing formal systems. However, unlike
LF, this logic cannot reflect dependencies between objects into types
and does not directly represent the relationship between formulas
and their proofs. While such correspondences can always be encoded by hand
through auxiliary predicate definitions, it is of interest to understand if
a systematic encoding is possible. A specific form to this
question is if Twelf specifications can be translated into $\lambda$Prolog 
programs,
allowing such specifications to be seen as \lprolog\ ``meta-programs.''
There are benefits to such a possibility: 
the convenience of writing specifications using dependent types can
be combined with the ability both to execute them via an efficient
$\lambda$Prolog implementation, and to reason about them
using logics and systems meant for analyzing \hhf\ descriptions
\cite{baelde10ijcar,gacek08ijcar,gacek08lics,miller05tocl}. 

\fussy

A partial answer to the question raised above has been
provided by Felty, who described a translation of 
LF specifications to \hhf\ formulas and then showed that 
LF derivations correspond exactly to \hhf\ derivations of
the translated LF judgment \cite{felty89phd,felty90cade}.
The focus on matching derivations allows Felty to assume the
existence of a complete LF judgment, and, in particular, of an LF object
in her translation. However, this assumption is inappropriate in our context,
given that we are interested in {\em constructing} proof terms that
show particular types are inhabited, \ie, in
{\em proof search} that plays a fundamental role in the logic
programming setting.
We therefore refine
the earlier mapping to remove this assumption and show that the
resulting translation preserves derivability in a sense relevant to the
logic programming interpretation; an important part of our proof is
showing how to extract an LF object satisfying a type from a
derivation constructed using the \hhf\ version of the specification.
Our first encoding may include redundant type-checking
judgments which obscure the translated specification
and can result in poor execution behavior.
We design conditions for eliminating some of these judgments,
resulting in an improved translation that corresponds
closely to the intention of the orginal LF specification. This part of
our work relies on an analysis of the structure of LF expressions and
also has relevance, for example, to providing compact representations
of proof terms. 
Finally, we demonstrate that the execution of the translated form
by means of the Teyjus implementation~\cite{teyjus.website} of
$\lambda$Prolog~\cite{nadathur88iclp}
provides an effective means for animating Twelf programs.

In the next two sections, we describe a relevant fragment
of the \hhf\ logic and the Twelf specification
language. Section~\ref{sec:translation} then presents our first
translation. In the following section, we
describe and exploit a property of LF expressions and type-checking to
refine the earlier translation, producing a more efficient and
transparent version. Section~\ref{sec:results} provides experimental
data towards 
supporting the use of this translation as a means for executing Twelf
programs. We conclude the paper with a discussion of related work and
possible future directions.
This work has been developed in~\cite{snow10masters}; we refer the reader to
that document for complete proofs and more detailed discussions.

\section{A Higher-Order Predicate Logic for Describing Computations}

The logic of \hhf\ formulas is based on an intuitionistic version of
Church's simple theory of types \cite{church40}. Both logics are
built over a typed form of the $\lambda$-calculus. The types are
constructed using $\rightarrow$, the infix, right associative function
type constructor, starting from a finite collection of atomic types
that includes $o$, the type of propositions, and at least one other
type.\footnote{Other, non-interpreted type constructors can be added
  but are not discussed here for simplicity.} 
We assume that we are given sets of variables and
constants, each with an associated type. The full collection of
(typed) terms  is generated from these by the usual abstraction
and (left associative) application operators. Terms that differ only
in the names of their bound variables are 
not distinguished. We further assume a notion of equality between
terms that is generated by $\beta$- and $\eta$-reduction. It is well-known
that every term has a unique normal form under these reduction
operations in this simply-typed setting. All terms are to be converted
into such a form prior to their consideration in 
any context. We write $t[s_1/x_1,\ldots,s_n/x_n]$ to denote the
result of simultaneously replacing the variables $x_1,\ldots,x_n$ with
the terms $s_1,\ldots,s_n$ in the term $t$, renaming bound variables
as needed to avoid accidental capture. This substitution operation is
defined only when $s_i$ and $x_i$ are of the same type for $1 \leq i \leq n$.  

We will use only a fragment of the full \hhf\ logic here; this
fragment still possesses the 
proof-theoretic properties that are fundamental to the logic
programming interpretation of the \hhf\ logic.
The constants from which terms are constructed are differentiated into 
{\it nonlogical} ones that constitute a {\it signature} and logical
ones. We do not permit $o$ to appear in the type of the
arguments of nonlogical constants and
variables. The logical constants are restricted to 
$\top$ of type $o$, $\supset$ of type $o \rightarrow o\rightarrow o$
that is written in the customary infix form and,
for each type $\alpha$, $\Pi$ of type
$(\alpha \rightarrow o) \rightarrow o$. 
$\Pi$ represents the universal quantifier as a function over sets. We
abbreviate $\Pi\app (\lambdax{x}{F})$ by 
$\forallx{x}{F}$. An {\it atomic formula}, denoted by $A$, is a term
of type $o$ of the form $p\app t_1\app \ldots\app 
t_n$ where $p$ is a nonlogical constant. The logic of interest is
characterized by two collections of terms called $G$- and
$D$-formulas that are defined mutually recursively by the following
syntax rules:
\begin{center}
\begin{tabular}{ccl}
   $G$ & $:=$ & $\top\ |\ A\ |\ D \supset G\ |\ \forallx{x}{G}$\\
   $D$ & $:=$ & $A\ |\ G \supset D\ |\ \forallx{x}{D}$\\
\end{tabular}
\end{center}
A specification or logic program is a finite collection of closed
$D$-formulas that are also called {\it program clauses} and a goal or
a query is a closed $G$-formula. 

\begin{figure*}  
\begin{center}
\begin{tabular}{ccc}
    \AxiomC{}
    \RightLabel{\topGoal}
    \UnaryInfC{$\sigsequent{\Sigma}{\Gamma}{\top}$}
    \DisplayProof

&

    \AxiomC{$\sigsequent{\Sigma}{\Gamma \cup\{D\}}{G}$}
    \RightLabel{\impGoal}
    \UnaryInfC{$\sigsequent{\Sigma}{\Gamma}{D \supset G}$}
    \DisplayProof

&

    \AxiomC{$c \notin \Sigma \quad \sigsequent{\Sigma \cup
        \{c\}}{\Gamma}{G[c/x]}$} 
    \RightLabel{\allGoal}
    \UnaryInfC{$\sigsequent{\Sigma}{\Gamma}{\forallx{x}{G}}$}
    \DisplayProof

\end{tabular}

\medskip

\begin{tabular}{cc}
    \AxiomC{$D\in \Gamma \quad \focusedsigseq{\Sigma}{\Gamma}{A}{D}$}
    \RightLabel{\atomicGoal}
    \UnaryInfC{$\sigsequent{\Sigma}{\Gamma}{A}$}
    \DisplayProof

&

    \AxiomC{}
    \RightLabel{\initial}
    \UnaryInfC{$\focusedsigseq{\Sigma}{\Gamma}{A}{A}$}
    \DisplayProof
\end{tabular}

\medskip
\begin{tabular}{cc}
    \AxiomC{$t\ \mbox{\rm is a}\ \Sigma\mbox{\rm -term} \quad
      \focusedsigseq{\Sigma}{\Gamma}{A}{D[t/x]}$}
    \RightLabel{\allClause}
    \UnaryInfC{$\focusedsigseq{\Sigma}{\Gamma}{A}{\forallx{x}{D}}$}
    \DisplayProof

&
    \AxiomC{$\sigsequent{\Sigma}{\Gamma}{G} \quad
      \focusedsigseq{\Sigma}{\Gamma}{A}{D}$}
    \RightLabel{\impClause}
    \UnaryInfC{$\focusedsigseq{\Sigma}{\Gamma}{A}{G \supset D}$}
    \DisplayProof
\end{tabular}
\end{center}
\caption{Derivation rules for the \hhf\ logic}
  \label{fig:hhfrules}
\end{figure*}

Computation corresponds to searching for a derivation  of a sequent of
the form 
$\sigsequent{\Sigma}{\Gamma}{G}$ where $\Sigma$ is the initial (language)
signature, $\Gamma$ is a logic program  and $G$ is a
goal. Figure~\ref{fig:hhfrules} presents the rules for constructing
such a derivation. Read in a proof search direction, the \allGoal\ rule leads
to an expansion of the 
signature in the sequent whose derivation is sought and the
\impGoal\ rule similarly
causes an addition to the logic program. The expression ``t is a
$\Sigma$-term'' in the \allClause\ rule means that $t$ is a closed term
all of whose nonlogical constants are contained in $\Sigma$. The
derivation rules manifest a goal-directed character: to find a 
derivation for $\sigsequent{\Sigma}{\Gamma}{G}$, we 
simplify $G$ based on its logical structure and then use the 
\atomicGoal\ rule to select a formula from the logic program for 
solving an atomic goal. Notice also that the \atomicGoal\ rule
initiates the consideration of a focused sequence of rules that is
similar to backchaining.\footnote{For the reader unfamiliar with such
  presentations, the expression
  $\focusedsigseq{\Sigma}{\Gamma}{A}{D}$ corresponds essentially to
  the selection of the program clause $D$ as the one to backchain
  on. This then leads to instantiations of universally quantified
  variables and to the solution of the ``body'' goals of the clause
  using the rules \allClause\ and \impClause, culminating eventually
  in solving the atomic goal by matching it with the head of the
  clause using the \initial\ rule.} In particular, if
the formula selected from $\Gamma$ has the structure 
\[(\forallx{x_1}{F_1 \supset \ldots \supset \forallx{x_n}{(F_n \supset
    A')\ldots)}}\]
then this sequence is equivalent to the rule
\begin{center}
    \AxiomC{$\sigsequent{\Sigma}{\Gamma}{F_1'}\qquad \ldots \qquad
      \sigsequent{\Sigma}{\Gamma}{F_n'}$}
    \RightLabel{\bcGoal}
    \UnaryInfC{$\sigsequent{\Sigma}{\Gamma}{A}$}
    \DisplayProof
\end{center}
which has the proviso that for some $\Sigma$-terms $t_1,\ldots,t_n$
that have the same types as $x_1,\ldots,x_n$, respectively, it is the
case that $A$ is equal to $A'[t_1/x_1,\ldots, t_n/x_n]$ and, for $1
\leq i\leq n$, $F_i'$ is equal to $F_i\subst{t_1 / x_1, \ldots, t_{i} /
  x_{i}}$.

The logic that we have described has been given an efficient
implementation in the {\it Teyjus} system
\cite{teyjus.website}. It is possible also to reason in sophisticated
ways about specifications that are constructed using it. To begin
with, the logic has strong meta-theoretic properties arising
from the fact that derivability in it corresponds exactly to
intuitionistic provability. Moreover, it is possible to construct logics
incorporating mechanisms such as induction to reason powerfully about
what does and does not follow from a given specification 
\cite{baelde08phd,gacek09phd,gacek08lics,miller05tocl}. In fact, systems such as
Abella \cite{gacek08ijcar} and Tac~\cite{baelde10ijcar} have been
constructed to provide computer support for such reasoning. 

\section{Logic Programming Using the Twelf Specification Language}
\label{sec:lf}
There are three categories of expressions in LF:
{\em kinds}, {\em types} or {\em type families} that are
classified by kinds and {\em objects} or {\em terms} that are
classified by types. We assume two denumerable sets of 
variables, one for objects and the other for types. We use $x$
and $y$ to denote object variables, $u$ and $v$ to denote type
variables and $w$ to denote either. Letting $K$ range over kinds, $A$
and $B$ over types, and $M$ and $N$ over object terms, the syntax of
LF expressions is given by the following rules:
\begin{center}
\begin{tabular}{ccl}
    $K$ & $:=$ & $\type\ |\ \typedpi{x}{A}{K}$ \\ 
    $A$ & $:=$ &
        $u\ |\ \typedpi{x}{A}{B}\ |\ \typedlambda{x}{A}{B}\ |\ A\app M $ \\ 
    $M$ & $:=$& $x\ |\ \typedlambda{x}{A}{M}\ |\ M\app N $ \\
\end{tabular}
\end{center}
Expressions of any of these kinds will be denoted by $P$ and $Q$. Here,
$\Pi$ and $\lambda$ are operators that associate a type with a 
variable and bind its free occurrences over the expression
after the period.  Terms that differ only in the
names of bound variables are identified. As with the
\hhf\ logic, $P[N_1/x_1,\ldots,N_n/x_n]$ denotes a simultaneous
substitution with renaming to avoid variable capture. We write $A
\rightarrow P$ for $\typedpi{x}{A}{P}$ when $x$ does not appear free
in $P$. We abbreviate $\typedpi{x_1}{A_1}{\ldots\typedpi{x_n}{A_n}{P}}$
by $\typedpis{x}{A}{P}$.

LF expressions are equipped with a notion of $\beta$-reduction defined
through the rule $(\typedlambda{x}{A}{P})\app N \rightarrow_\beta
P[N/x]$. All LF expressions that are well-formed in the sense
formalized below normalize strongly under this reduction
relation \cite{harper93jacm}. Moreover any well-typed expression $P$
has a unique normal form up to changes in bound variable names. We
denote this normal form by $P^\beta$.  

The type correctness of LF expressions is assessed relative to
contexts that are finite collections of assignments of types and kinds
to variables. Formally, contexts, denoted by $\Gamma$, are given by the
rule 
\begin{center}
\begin{tabular}{ccl}
$\Gamma$ & $:=$ &
  $\emptyctx\ \vert\ \Gamma,\oftype{u}{K}\ \vert\ \Gamma,\oftype{x}{A}$\\
\end{tabular}
\end{center}
Here, $\emptyctx$ denotes the empty collection. We write 
$dom(\Gamma)$ to denote the variables with assignments in
$\Gamma$. We are concerned with assertions of the following four forms:
\begin{center}
$\lfprove{}{\Gamma\ \ctx} \qquad \lfprove{\Gamma}{K \ \kind} \qquad
\lfprove{\Gamma}{\oftype{A}{K}} \qquad \lfprove{\Gamma}{\oftype{M}{A}}$ 
\end{center}
The first assertion signifies that $\Gamma$ is a
well-formed context. The remaining assertions mean
respectively that, relative to a (well-formed) context $\Gamma$, $K$
is a well-formed kind, $A$ is a well-formed type of kind $K$ and $M$
is a well-formed object of type $A$. 
Figure~\ref{fig:lf-rules} presents the rules for deriving such
assertions. Notice that for a context to be well-formed it must not
contain multiple assignments to the same variable. To adhere to this
requirement, bound variable renaming may be entailed in the use of the
\pikind, \pifam, \absfam\ and \absobj\ rules. The inference rules
allow for the derivation of an assertion of the form
$\lfprove{\Gamma}{\oftype{M}{A}}$ only when $A$ is in normal form. To
verify such an assertion when $A$ is not in normal form, we first
derive $\lfprove{\Gamma}{\oftype{A}{\type}}$ and then verify
$\lfprove{\Gamma}{\oftype{M}{A^\beta}}$. A similar observation applies
to $\lfprove{\Gamma}{\oftype{A}{K}}$. 

\begin{figure*}
\begin{center}
   \AxiomC{}
   \RightLabel{\nullctx}
   \UnaryInfC{$\lfprove{}{\emptyctx\ \ctx}$}
   \DisplayProof

\medskip

   \AxiomC{$\lfprove{\Gamma}{K\ \kind} \quad
     \lfprove{}{\Gamma\ \ctx}\quad u \notin dom(\Gamma)$} 
   \RightLabel{\kindctx}
   \UnaryInfC{$\lfprove{}{\Gamma, u : K\ \ctx}$}
   \DisplayProof

\medskip

   \AxiomC{$\lfprove{\Gamma}{\oftype{A}{\type}} \quad
     \lfprove{}{\Gamma\ \ctx}\quad x \notin dom(\Gamma)$}
   \RightLabel{\typectx}
   \UnaryInfC{$\lfprove{}{\Gamma, x : A\ \ctx}$}
   \DisplayProof

\medskip

\begin{tabular}{cc}
   \AxiomC{$\lfprove{}{\Gamma\ \ctx}$}
   \RightLabel{\typekind}
   \UnaryInfC{$\lfprove{\Gamma}{\type\ \kind}$}
   \DisplayProof

&
   \AxiomC{$\lfprove{\Gamma}{\oftype{A}{\type}} \quad
     \lfprove{\Gamma,\oftype{x}{A}}{K\ \kind}$}
   \RightLabel{\pikind}
   \UnaryInfC{$\lfprove{\Gamma}{\typedpi{x}{A}{K}\ \kind}$}
   \DisplayProof
\end{tabular}

\medskip
\begin{tabular}{cc}
   \AxiomC{$\lfprove{}{\Gamma\ \ctx} \quad \oftype{u}{K} \in \Gamma$}
   \RightLabel{\varfam}
   \UnaryInfC{$\lfprove{\Gamma}{\oftype{u}{K^\beta}}$}
   \DisplayProof

&
   \AxiomC{$\lfprove{}{\Gamma\ \ctx} \quad \oftype{x}{A} \in \Gamma$}
   \RightLabel{\varobj}
   \UnaryInfC{$\lfprove{\Gamma}{\oftype{x}{A^\beta}}$}
   \DisplayProof
\end{tabular}

\medskip

   \AxiomC{$\lfprove{\Gamma}{\oftype{A}{\type}} \quad \lfprove{\Gamma,
       \oftype{x}{A}}{\oftype{B}{\type}}$}
   \RightLabel{\pifam}
   \UnaryInfC{$\lfprove{\Gamma}{\oftype{(\typedpi{x}{A}{B})}{\type}}$}
   \DisplayProof

\medskip
\begin{tabular}{cc}
   \AxiomC{$\lfprove{\Gamma}{\oftype{A}{\type}} \quad
     \lfprove{\Gamma,\oftype{x}{A}}{\oftype{B}{K}}$}
   \RightLabel{\absfam}
   \UnaryInfC{$\lfprove{\Gamma}{\oftype{(\typedlambda{x}{A}{B})}{(\typedpi{x}{A^\beta}{K})}}$}
   \DisplayProof

&

   \AxiomC{$\lfprove{\Gamma}{\oftype{A}{\typedpi{x}{B}{K}}} \quad
     \lfprove{\Gamma}{\oftype{M}{B}}$}
   \RightLabel{\appfam}
   \UnaryInfC{$\lfprove{\Gamma}{\oftype{(A\app M)}{(K[M/x])^\beta}}$}
   \DisplayProof
\end{tabular}

\medskip
\begin{tabular}{cc}
   \AxiomC{$\lfprove{\Gamma}{\oftype{A}{\type}} \quad
     \lfprove{\Gamma,\oftype{x}{A}}{\oftype{M}{B}}$}
   \RightLabel{\absobj}
   \UnaryInfC{$\lfprove{\Gamma}{\oftype{(\typedlambda{x}{A}{M})}{(\typedpi{x}{A^\beta}{B})}}$}
   \DisplayProof

&

   \AxiomC{$\lfprove{\Gamma}{\oftype{M}{\typedpi{x}{A}{B}}} \quad
     \lfprove{\Gamma}{\oftype{N}{A}}$}
   \RightLabel{\appobj}
   \UnaryInfC{$\lfprove{\Gamma}{\oftype{(M\app N)}{(B[N/x])^\beta}}$}
   \DisplayProof
\end{tabular}
\end{center}
  \caption{Rules for Inferring LF Assertions}
  \label{fig:lf-rules}
\end{figure*}

A variable $w$ that appears in an LF expression $P$ that is
well-formed with respect to a context $\Gamma$ has
a kind or type of kind \type\ associated with it through either an
assignment in $\Gamma$ or a binding operator. Moreover, the normal form
of this kind or type must have a prefix of $\Pi$s. If the length of
this prefix is $n$, then an occurrence of $w$ is {\it fully applied}
if it appears in a subterm of the form $w\app M_1\app \ldots\app M_n$. 
Further, $P$ is {\em canonical} with 
respect to $\Gamma$ if it is in normal form and if every
variable occurrence in it is 
fully applied. A well-formed context $\Gamma$ is
canonical if the type or kind it assigns to each variable is canonical
relative to $\Gamma$. A well-formed type of the form $u\app M_1 \app
\ldots\app M_n$ that is fully applied is called a {\em base type}. The
LF system admits a notion of $\eta$-expansion using which any
well-formed expression can be converted into a canonical form. 

In later sections we shall consider LF derivations in which all
expressions in the end assertion are in normal form. Notice
that every expression in the entire derivation must then
also be in such a form. This in turn means that in judgments of
the forms $\oftype{(\typedlambda{x}{A}{B})}{(\typedpi{x}{A'}{K})}$ and 
$\oftype{(\typedlambda{x}{A}{M})}{(\typedpi{x}{A'}{B})}$ it must be
the case that $A$ and $A'$ are identical. Finally, normalization need
not be considered in the use of the \varfam\ and \varobj\ rules.

The following ``transitivity'' property for LF derivations that
follows easily from the results in \cite{harper93jacm} will be
useful later; here $\alpha$ stands for any judgment,
and substitution and normalization over
$\alpha$ and $\Gamma$ corresponds to distributing these operations to
the expressions appearing in them.

\begin{proposition}[Substitution]
  \label{prop:lf-substitution}
  Let $\Gamma_1$, $\Gamma_2$ be canonical contexts,
  and $A$ be a type in canonical form.
  If $\Gamma_1 \vdash M : A$ has a derivation, and
  $\Gamma_1, x : A, \Gamma_2 \vdash \alpha$ has a derivation, then
  $\Gamma_1, (\Gamma_2[M / x])^\beta \vdash (\alpha[M / x])^\beta$
  has a derivation as well.
\end{proposition}

Additionally we will use a second property of LF derivations,
which follows from Proposition~\ref{prop:lf-substitution}.

\begin{proposition}[Renaming]
  \label{prop:lf-renaming}
  Let $P$ be a canonical type or kind, 
  $\Gamma = \Gamma_1, \oftype{x}{P}, \Gamma_2$ be a canonical context,
  and $\alpha$ a canonical judgment.  Let $y$ be a variable
  not bound in $\Gamma$, and not occurring in $\alpha$.  Then
  $\lfprove{\Gamma_1, \oftype{x}{P}, \Gamma_2}{\alpha}$ has
  a derivation if and only if
  $\lfprove{\Gamma_1, \oftype{y}{P}, \Gamma_2 \subst{y / x}}{\alpha \subst{y / x}}$
  has one.
\end{proposition}

The logic programming interpretation of LF is based on viewing 
types as formulas. More specifically, a specification or program in
this setting is given by a context. This starting context, also called
a {\em signature}, essentially describes the vocabulary for
constructing types and asserts the existence of particular inhabitants
for some of these types. Against this backdrop, questions can be asked
about the existence of inhabitants for certain other types. Formally,
this amounts to asking if an assertion of the form
$\lfprove{\Gamma}{\oftype{M}{A}}$ has a derivation. However, the
object $M$ is left unspecified---it is to be extracted
from a successful derivation. Thus, the search for a derivation of 
the assertion is driven by the structure of $A$ and the types
available from the context.  

A concrete illustration of the paradigm is useful for later
discussions.\footnote{The example of appending lists has been chosen
  here for its conciseness and because it allows for an easy connection
  with more traditional forms of logic programming. The primary
  application domain of Twelf is in specifying (and reasoning about)
  formal systems such as evaluators and interpreters for languages,
  type assignment calculi and proof systems. This orientation informs
  the choice of benchmarks used in Section~\ref{sec:results}.}
Consider a signature or  context $\Gamma$ comprising the 
following assignments in sequence: 
\begin{tabbing}
\qquad\=\qquad\quad\=\qquad\=\kill
\>$\oftype{nat}{\type}$\\
\>$\oftype{z}{nat}$\\
\>$\oftype{s}{nat \rightarrow nat}$\\
\>$\oftype{list}{\type}$\\
\>$\oftype{nil}{list}$\\
\>$\oftype{cons}{nat \rightarrow list \rightarrow list}$\\ 
\>$\oftype{append}{list \rightarrow list \rightarrow list \rightarrow \type}$\\
\>$\oftype{appNil}{\typedpi{K}{list}{append\app nil\app K\app K}}$\\
\>$\oftype{appCons}{\typedpi{X}{nat}{\typedpi{L}{list}{\typedpi{K}{list}{\typedpi{M}{list}{}}}}}$\\
\>\>$(append \app L \app K \app M) \rightarrow$\\
\>\>\>$(append\app (cons\app X\app L)\app K\app (cons\app X\app M))$
\end{tabbing}
We can ask if there is some term $M$ such that the judgment 
\begin{tabbing}
\qquad\=$\lfprove{\Gamma}{\oftype{M}{append\app }}$\=\kill
\>$\lfprove{\Gamma}{\oftype{M}{append\app (cons\app z\app nil)}}$\\
\>\>$(cons\app (s\app z)\app nil)$\\
\>\>$(cons \app z\app (cons \app (s\app z)\app nil))$
\end{tabbing}
is derivable. Assuming that $\Gamma$ is given by the ambient
environment, such a query can be posed in Twelf simply by presenting the
type expression.
The logic programming interpreter of Twelf will find that the proof term 
\begin{tabbing}
\qquad\=$(appCons\app z\app nil \app $\=\kill
\>$(appCons\app z\app nil \app (cons\app (s\app z)\app nil)$\\
\>\>$(cons\app (s\app z)\app nil)$\\
\>\>$(appNil\ (cons\app (s\app z)\app nil)))$
\end{tabbing}
inhabits this type and hence will succeed on the query. In reaching
this conclusion, the interpreter will use the types involving $append$
that are present in $\Gamma$. Further it will do this in a way that
bears a close resemblance to the use of clauses in a Prolog-like
setting, interpreting $\Pi$ like a universal quantifier and
$\rightarrow$ like an implication. 

The simple example we have considered here will suffice to illustrate
most of the later ideas in this paper but it does not bring out the
richness of dependent types in specifications. We leave this
demonstration to the many discussions already in the literature. We
also note that Twelf has many additional features like allowing $\Pi$
quantification in types to be left implicit and permitting
instantiatable variables in queries whose values are to be found through
unification. While these aspects are treated in our implementation, to
keep the theoretical discussions focused, we shall assume that the only
capability that is to be emulated is that of determining the
derivability of an assertion of the form
$\lfprove{\Gamma}{\oftype{M}{A}}$ in which $\Gamma$ and $A$ are in
canonical form (and $M$ is left unspecified). This assumption is easily
justified: these will be 
``type-checked'' prior to conducting a search and the Twelf system
assumes equality under $\eta$-conversion.

\section{From Twelf Specifications to Predicate Formulas}
\label{sec:translation}

\begin{figure*}
  \centering
   \[ \begin{array}{c}
   \phi(A) := \lfobject \ \text{when $A$ is a base type} \\
   \phi(\typedpi{x}{A}{P}) := \phi(A)\ra \phi(P)
   \quad\quad
   \phi(\type) := \lftype
   \end{array} \]
\[\begin{array}{c}
    \encTerm{u\ M_1 \ldots M_n} := u \ \encTerm{M_1}\ldots\encTerm{M_n} \\
\quad\quad
    \encTerm{x\ M_1 \ldots M_n} := x \ \encTerm{M_1}\ldots \encTerm{M_n}
\quad\quad
    \encTerm{\typedlambda{x}{A}{M}} := \lambda^{\phi(A)} x. \encTerm{M}
\end{array} \]
  \begin{align*}
    \enc{\typedpi{x}{A}{B}} :=&\
      \lambda M.~
      \forall x.~ (\enc{A}\app x) \supset (\enc{B}\app (M\app x)) \\
    \enc{A} :=&\
      \lambda M.~
      hastype \ M\ \encTerm{A}
      \ \text{where $A$ is a base type}
  \end{align*}
  \caption{Encoding of types, objects, and
     simplified translation of LF judgments to \hhf}
  \label{fig:simplified-translation}
\end{figure*}

\begin{figure*}
\centering
\begin{tabbing}
\qquad\qquad\qquad\=\qquad\=\kill
\>$hastype \ z \ nat$ \\
\>$\forall n.~ hastype \ n \ nat \supset hastype \ (s \ n) \ nat$ \\
\>$hastype \ nil \ list$ \\
\>$\forall n.~
       hastype \ n \ nat \supset \forall l.~ \hastype \ l \ list \supset
       hastype \ (cons \ n \ l) \ list$ \\
\>$\forall l.~ hastype \ l \ list \supset
      hastype \ (appNil \ l) \ (append \ nil \ l \ l)$ \\
\>$\forall x.~ hastype \ x \ nat \supset
   \forall l.~ hastype \ l \ list \supset
   \forall k.~ hastype \ k \ list \supset
   \forall m.~ hastype \ m \ list \supset$ \\
\>\>$\forall a.~ hastype \ a \ (append \ l \ k \ m) \supset hastype \ (appCons \ x \ l \ k \ m \ a) \ (append \ (cons \ x \ l) \ k \ (cons \ x \ m))$
\end{tabbing}
\caption{Simple translation of the LF specification for $append$}
\label{fig:simplified-append-translation}
\end{figure*}

Felty has previously shown how to translate LF specifications and
judgments into \hhf\ formulas \cite{felty89phd, felty90cade}. Her
translation proceeds in two steps. First, she describes a coarse
mapping of LF expressions into (simply typed) $\lambda$-terms. This
mapping loses information about dependencies in types and kinds and
also does not reflect the correspondences between objects and types
and types and kinds. These relationships are encoded later
through binary predicates over $\lambda$-terms. 

The general structure of Felty's translation is applicable in the
context of interest to us. However, the details of her mapping do not
quite fit our needs because of her focus on {\it derivations} in the
LF and \hhf\ logics. One manifestation of this is that her translation
is not based exclusively on types, but assumes also the
availability of the objects they are intended to qualify. This is
not acceptable in the context of proof search where the task is precisely
to determine the existence of those objects: we need a translation that is
only based on the type, and which can be applied to an \hhf\ metavariable
to correspond to an LF query whose object is left unspecified as a
metavariable.
Second, the correctness result only states an equivalence between LF
derivability and \hhf\ derivability for {\it known} LF assertions, 
and does not consider, for example, whether it is possible for
non-canonical or ill-formed objects to be produced in the course of
searching for proofs from the \hhf\ specification.
In contrast, our completeness result will guarantee that after
running a query with a metavariable standing for the (encoding of the)
object, the only possible instantiations of that metavariable
are actual encodings of terms.

The first step towards producing a translation into \hhf\ that can be
used to interpret Twelf specifications is to adapt Felty's translation
in a way that makes it acceptable in logic programming discussions. 
Our translation shall only account for judgments of the form
$\lfprove{\Gamma}{\oftype{M}{A}}$ since these are the only ones of
interest in the logic programming setting described in the previous
section. The adequacy of this restriction actually relies on an
auxiliary, easily verified, fact: 
if $\lfprove{\Gamma}{\oftype{A}{\type}}$ is known to have a
derivation and the last rule in a purported derivation of
$\lfprove{\Gamma}{\oftype{M}{A}}$ is an \absobj, then the left premise
for the latter derivation must have a proof and hence does not need to
be encoded by the translation. 

Our translation is presented in
Figure~\ref{fig:simplified-translation}. This
translation first encodes LF objects and types in \hhf\ terms by
dropping a lot of typing information; as mentioned already, this
information will be recovered later in  
the encoding of LF judgments. Under this translation, an object (type)
of type (kind) $P$ is 
represented by an \hhf\ term of simple type $\phi(P)$, built from 
the atomic types $\lftype$ and $\lfobject$. The encoding of an object or
base type $Q$ is then given by $\encTerm{Q}$; note that in the process
we assume a reuse of (LF) variable names with an appropriate type as
part of the corresponding \hhf\ signature. 
As an example, the LF signature at the end of the last section leads
to the following \hhf\ signature:
\begin{tabbing}
\qquad\=\qquad\quad\=\qquad\=\kill
\>$\oftype{nat}{\lftype}$\\
\>$\oftype{z}{\lfobject}$\\
\>$\oftype{s}{\lfobject \rightarrow \lfobject}$\\
\>$\oftype{list}{\lftype}$\\
\>$\oftype{nil}{\lfobject}$\\
\>$\oftype{cons}{\lfobject \rightarrow \lfobject \rightarrow \lfobject}$\\ 
\>$\oftype{append}{\lfobject \rightarrow \lfobject \rightarrow \lfobject 
                       \rightarrow \lftype}$\\
\>$\oftype{appNil}{\lfobject \ra \lfobject}$\\
\>$\oftype{appCons}{\lfobject\ra\lfobject\ra}$\\
\>\>\>$\lfobject\ra\lfobject\ra\lfobject\ra\lfobject$
\end{tabbing}
Further, the LF type $append \ nil \ nil \ nil$ gets translated
to the same term in \hhf, where it has type $\lftype$.
This translation behaves well with respect to
substitution and $\beta$-conversion,
and is injective for objects (types) of the same type (kind).
Finally, we take up the translation of LF type assignments and
judgments in the last two clauses in
Figure~\ref{fig:simplified-translation}. 
To emphasize reliance only on the structure of types, these clauses
describe explicitly only the translation of an LF type
$A$. Such a type is mapped into an \hhf\ predicate denoted by
$\enc{A}$ that, intuitively, codifies the property of being a 
translation of an LF object of type $A$. 
This translation is defined on all canonical types and 
uses the \hhf\ predicate $hastype$ of type
$\lfobject \ra \lftype \ra o$.
If $A$ is a base type, $\enc{\typedpi{x_1}{B_1}{\ldots\typedpi{x_n}{B_n}{A}}}$
has type $\tau \ra o$ where
$\tau$ is $\lfobject \ra \ldots \ra \lfobject \ra \lfobject$ 
with $n$ negative occurrences of $\lfobject$. Once the translation of
LF types is in place, we define $\enc{\oftype{M}{A}}$ derivatively to
be $(\enc{A}\app \encTerm{M})$.  

Twelf specifications are encoded by dropping all kind assignments
and translating each type assignment they contain.
As an example, the Twelf specification
of $append$ translates into the clauses in
Figure~\ref{fig:simplified-append-translation}. From these clauses, we
can, for example, derive the goal $hastype \ (cons \ (s \ z) \ nil)
\ list$ and we could search for terms $X$ satisfying the goal
\begin{tabbing}
\qquad\=$hastype\ X\ (append\ $\=\kill
\>$hastype \ X \ (append \ (cons \ z \ nil)$\\
\>\>$(cons \ (s \ z) \ nil)$\\
\>\>$(cons \ z \ (cons \ (s \ z) \ nil)))$.
\end{tabbing}

Let $\Gamma'$ be the translation of an LF context $\Gamma$ and
$\alpha'$ be the translation of the LF judgment $\alpha$. These
translations are based on an implicit \hhf\ signature $\Sigma$. In the
case that all the free variables in $\alpha$ belong to $dom(\Gamma)$,
then, in fact, $\Sigma$ consists of an isomorphic copy of the symbols
in $dom(\Gamma)$. Henceforth, we shall assume $\Sigma$ to be just
such an \hhf\ signature and we shall write $\iprove{\Gamma'}{\alpha'}$ 
as a shorthand for $\iprove{\Sigma; \Gamma'}{\alpha'}$. The
correctness of the (simple) translation is then the content of the
following theorem. 

\begin{theorem'}
  \label{theorem:simplified-translation-correctness}
  Let $\Gamma$ be a well-formed canonical LF context and let 
  $A$ be a canonical LF type such that
  $\lfprove{\Gamma}{\oftype{A}{\type}}$ has a derivation. 
  If $\lfprove{\Gamma}{\oftype{M}{A}}$ has a derivation for a canonical
  object $M$, then there is a derivation of
  $\iprove{\enc{\Gamma}}{\enc{\oftype{M}{A}}}$.
  Conversely, if $\iprove{\enc{\Gamma}}{(\enc{A}\app M)}$ has a derivation
  for any   \hhf\ term $M$ of appropriate type, then there is a
  canonical LF  object $M'$ such that   $M = \encTerm{M'}$ and
  $\lfprove{\Gamma}{\oftype{M'}{A}}$ has a derivation. 
\end{theorem'}

\paragraph{\it Proof outline}
Completeness can be proved by a simple induction
on the LF derivation, building an \hhf\ derivation that mimics its
structure. 
Soundness is more involved: we proceed by induction
on the \hhf\ derivation, gradually recovering the structure of
$M'$, maintaining the derivability of
$\lfprove{\Gamma}{\oftype{A}{\type}}$ that  allows
us to build an LF derivation even in the case that \absobj\ was the
last rule used. The detailed proof is presented in Appendix~\ref{appendix:proofs}.

\medskip
The simple translation presented in this section cannot be the basis
of a practical implementation of 
logic programming in LF.  Proof search using a program it produces 
may involve repeatedly proving goals of the form $hastype \ M \ A$ for
(encodings of) the same object $M$ and type $A$. 
This can be seen from the example in 
Figure~\ref{fig:simplified-append-translation}: at every step of deriving
an instance of $append$, the lists must be checked to be well-typed,
which artificially introduces a quadratic complexity. An important
point to note, however, is that this redundancy in ``type-checking''
is not easily detectable from the \hhf\ program that is
generated. Rather, it must be determined, and shown to be safely
eliminable, based on deeper properties of LF terms. It is this issue
that we take up in the next section.

\section{An Improved Translation of Twelf Specifications}
\label{sec:optimization}

\begin{figure}
\centering
\renewcommand{\arraystretch}{3}
\begin{tabular}{c}
    \AxiomC{$\termUV{\Gamma; \cdot; x}{A_i}$ for some $A_i$}
    \RightLabel{\ruvappt}
    \UnaryInfC{$\formulaUV{\Gamma; x}{c \vec{A}}$}
\DisplayProof
\\
  \AxiomC{$y_i \in \delta$ for each $y_i \quad\ y_i$ distinct}
  \RightLabel{\ruvinito}
  \UnaryInfC{$\termUV{\Gamma; \delta; x}{x \app \vec{y}}$}
\DisplayProof
\\
  \AxiomC{$\formulaUV{\Gamma, y; x}{B}$}
  \RightLabel{\ruvpit}
  \UnaryInfC{$\formulaUV{\Gamma; x}{\typedpi{y}{A}{B}}$}
\DisplayProof
\\
  \AxiomC{$y \notin \Gamma$ and $\termUV{\Gamma; \delta; x}{M_i}$ for some $i$}
  \RightLabel{\ruvappo}
  \UnaryInfC{$\termUV{\Gamma; \delta; x}{y\ \vec{M}}$}
\DisplayProof
\\
  \AxiomC{$\termUV{\Gamma; \delta, y; x}{M}$}
  \RightLabel{\ruvabso}
  \UnaryInfC{$\termUV{\Gamma; \delta; x}{\typedlambda{y}{A}{M}}$}
\DisplayProof
\end{tabular}
  \caption{Rigidly occurring variables in types and objects}
  \label{fig:ruvs}
\end{figure}

In order to make the translation of LF specifications into \hhf\ practical
from an implementation standpoint, we make two optimizations.

The first, and main, optimization exploits the fact that we are considering
derivations of the form $\lfprove{\Gamma}{\oftype{M}{A}}$
where $\Gamma$ and $A$ have already been type-checked.
For example, we may be wanting to determine whether the LF type 
\[append \app (cons \app z \app nil) \app nil \app (cons \app z \app nil)\]
is inhabited. Before attempting to do this, we would have already
determined that $append \app (cons \app z \app nil) \app nil \app
(cons \app z \app nil)$
is a valid type, which means, for instance, that we would have checked
that $(cons \app z \app nil)$ is a valid object of type $list$. Therefore,
there is no need to show again that $(cons \app z \app nil)$ has this
property in the course of searching for an inhabitant of the displayed
type. Our optimized translation takes advantage of this kind
of observation by statically removing some run-time checking
from the translation of LF typing.
More specifically, our optimization is
based on the following idea. Suppose we can determine that, for a
particular $i$, $t_i$ must always appear in the type
$(A\subst{t_1/x_1,\ldots,t_n/x_n})^\beta$. Then the translation of the
type $\typedpi{x_1}{B_1}{\ldots\typedpi{x_n}{B_n}{A}}$ does not need
to include explicit type-checking over the instantiation of
$x_i$.
We characterize some of these cases by using the notion of 
a \emph{rigid occurrence of $x_i$ in $A$}
that is expressed formally through the judgment
$\formulaUV{\vec{x} ; x_i}{A}$
defined by the rules in Figure~\ref{fig:ruvs};
the rules \ruvappt\ and \ruvpit\ in this figure act on LF types,
and the rules \ruvinito, \ruvappo, and \ruvabso\ act on LF objects.
We shall allow type checking over instantiations of rigid variables
to be eliminated from the simple translation. By doing so, we shall
both reap an efficiency benefit and also make the result of translation
correspond more closely to the original LF type.  

The second optimization is more transparent, not depending on deep
properties of dependent types. The essential observation is the
following. Instead of producing predicates of the form
\begin{tabbing}
\qquad\=\kill
\>$hastype \app X \app (append \app L \app K \app M)$
\end{tabbing}
and $hastype \app L \app list$, we can specialize them to $append \app
X \app L \app K \app M$ and $list \app L$. 
This results in a \hhf\ program that is much clearer,
and more closely related to the original LF specification.
Moreover,
this  simple transformation can also lead to better performance in a
logic programming setting because it allows for the exploitation of a
common optimization, namely, the indexing on a predicate name that
speeds up the determination of candidate clauses on which to backchain.

The improved translation that uses these two ideas is presented on 
Figure~\ref{fig:optimized-translation}.
The $\encExtP{\bullet}{\Gamma}$ translation is used on type assignments
appearing negatively (notably context items) and $\encExtN{\bullet}$ on
positive typing judgments (notably the conclusion of LF assertions).
As before, that translation is entirely guided by the type,
and defined for all canonical types.
We shall use the notation $\encExtN{\oftype{M}{A}}$ for
$(\encExtN{A} \encTerm{M})$, and define $\encExtP{\Gamma}{}$ as the
result of applying $\encExtP{\bullet}{\cdot}$ to each context item, dropping
kind assignments.  Note that instead of replacing unnecessary typing
judgments with $\top$ we could simply elide them all together;
we use $\top$ as a placeholder because it simplifies later proofs.
This translation is illustrated by its application to the example
Twelf specification considered in Section~\ref{sec:lf} that yields the
clauses shown in Figure~\ref{fig:optimized-append}. These clauses
should be contrasted with the ones in
Figure~\ref{fig:simplified-append-translation} that are produced by
the earlier, naive translation.

\begin{figure*}
  \centering
  \begin{align*}
    \encExtP{\typedpi{x}{A}{B}}{\Gamma} :=&\
      \begin{cases}
        \lambda M.~ \forall x.~ \top \supset \encExtP{B}{\Gamma, x}(M \app x)
          & \text{if}\ \formulaUV{\Gamma; x}{B} \\
        \lambda M.~ \forall x.~
            \encExtN{A}(x) \supset \encExtP{B}{\Gamma, x}(M \app x)
          & \text{otherwise}
      \end{cases} \\
    \encExtP{u \vec{N}}{\Gamma} :=&\ \lambda M.~ u \app M \app \vec{\encTerm{N}}
    \\
    \encExtN{\typedpi{x}{A}{B}} :=&\
      \lambda M.~ \forall x.~
         \encExtP{A}{\cdot}(x) \supset \encExtN{B}(M \app x) \\
    \encExtN{u \vec{N}} :=&\ \lambda M.~ u \app M \app \vec{\encTerm{N}}
  \end{align*}
  \caption{Optimized translation of LF specifications and judgments to \hhf}
  \label{fig:optimized-translation}
\end{figure*}

\begin{figure*}
\begin{tabbing}
\qquad\qquad\qquad\qquad\=\qquad\qquad\=\kill
\>$nat \app z$ \\
\>$\forall n.~ nat \app n \supset nat \app (s \app n)$ \\
\>$list \app nil$ \\
\>$\forall n.~
       nat \app n \supset  \forall l.~ list \app l \supset
       list \app (cons \app n \app l)$ \\
\>$\forall l.~ \top \supset
      append \app (appNil \app l) \app nil \app l \app l$ \\
\>$\forall x.~ \top \supset
   \forall l.~ \top \supset
   \forall k.~ \top \supset
   \forall m.~ \top \supset$ \\
\>\>$\forall a.~ append \app a \app l \app k \app m \supset append \app (appCons \app x \app l \app k \app m \app a) \app (cons \app x \app l) \app k \app (cons \app x \app m)$
\end{tabbing}
\caption{Optimized translation of the LF specification for $append$}
\label{fig:optimized-append}
\end{figure*}

We shall now establish the correctness of the optimized translation.
We first prove a fundamental lemma concerning rigidly occurring variables,
that is in fact an observation about LF:
for an LF base type $A$,
if we have derivations of
\[ \begin{array}{l}
\lfprove{\Gamma}{
       \oftype{\typedpi{x_1}{B_1}{\ldots \typedpi{x_n}{B_n}{A}}}{\type}}
       \qquad \mbox{\rm and}
\\
\lfprove{\Gamma}{\oftype{A \subst{t_1/x_1\ldots t_n/x_n}}{\type}}
\end{array} \]
and there is a rigid occurrence of $x_i$ in $A$, \ie,
$\formulaUV{\vec{x}; x_i}{A}$ has a derivation, 
then
$\lfprove{\Gamma}{\oftype{t_i}{B_i \subst{t_1/x_1\ldots t_{i-1}/x_{i-1}}}}$
has a derivation.
The idea of the proof is as follows.
The judgment $\formulaUV{\vec{x}; x_i}{A}$
gives a path in $A$ that leads to $x_i$, and this path can never be
erased by the considered substitution; following this path
simultaneously in the two LF derivations, one eventually finds
on one side a derivation of $\lfprove{\Gamma}{\oftype{x_i}{B_i}}$
and on the other side the expected derivation of
$\lfprove{\Gamma}{\oftype{t_i}{B_i \subst{t_1 / x_1, \ldots, t_{i-1} / x_{i-1}}}}$.

In order to be able to use this observation in our correctness
argument, we formulate a stronger, rather technical lemma that deals
directly with encoded types that are the result of instantiations
of (a priori) arbitrary \hhf\ terms, and ensures
that discovered \hhf\ terms are in fact encodings of LF objects.
These technical details concerning encodings are tedious but shallow,
and the essential structure of the proof follows the lines sketched
above. 

\begin{definition}
  Let $\vec{t}$ be a vector of \hhf\ terms, and $\vec{x}$ a vector of
  variables of the same length. If $M$ and $N$ are LF objects, then
  we write $(M \sim N)\subst{t_1/x_1\ldots t_n/x_n}$
  when \[\encTerm{M} = \encTerm{N} \subst{t_1/x_1\ldots t_n/x_n}.\]
  For LF types $A$ and $B$, we write $(A \sim B)\subst{t_1/x_1\ldots t_n/x_n}$
  when the two types are equal up to
  $(\bullet \sim \bullet)\subst{t_1/x_1\ldots t_n/x_n}$ on objects within.
  Finally we extend this notion to contexts
  of the same length by pushing it down to the types bound by the context.
  We shall omit $\vec{t}$ and $\vec{x}$ when they are obvious from the context,
  simply writing $\encEq{P}{Q}$.
\end{definition}

\begin{lemma'}
  \label{lemma:rigid-variables}
  Let $\vec{t}$ be a vector of \hhf\ terms,
  $\vec{x}$ a vector of variables, and $\vec{B}$ of canonical LF types,
  all of same length,
  such that $t_j = \encTerm{t'_j}$ for $j < i$.
  Let $\Gamma_0 = x_1 : B_1, \ldots, x_n : B_n$.
\begin{enumerate}
\item
  Let $\Gamma$ and $\Delta$ be LF contexts,
  $M$ an LF object and $A$ a type,
  all being assumed canonical.
  Let $\delta$ be $dom(\Delta)$.
  Suppose that there are derivations of
  $\termUV{\vec{x}; \delta; x_i}{M}$ and
  $\lfprove{\Gamma, \Gamma_0, \Delta}{\oftype{M}{A}}$ and
  $\lfprove{\Gamma, \Delta'}{\oftype{M'}{A'}}$,
  with $\encEq{A'}{A}$, $\encEq{M'}{M}$ and $\encEq{\Delta'}{\Delta}$.
  Then $t_i$ is of the form $\encTerm{t'_i}$ and there is a derivation of
  $\lfprove{\Gamma}{
      \oftype{t'_i}{B_i \subst{t'_1 / x_1, \ldots, t'_{i-1} / x_{i-1}}}}$.
\item
  Let $\typedpis{x}{B}{A}$ be a canonical type, where $A$ is a base type.
  Suppose that
  $\lfprove{\Gamma}{\oftype{\typedpis{x}{B}{A}}{\type}}$ and
  $\formulaUV{\vec{x}; x_i}{A}$ have derivations. Further, for some
  $A'$ such that $\encEq{A'}{A}$, suppose that
  $\lfprove{\Gamma}{\oftype{A'}{\type}}$ has a derivation. 
  Then $t_i = \encTerm{t'_i}$ and there is a derivation of
  $\lfprove{\Gamma}{
       \oftype{t'_i}{B_i \subst{t'_1 / x_1, \ldots, t'_{i-1} / x_{i-1}}}}$.
\end{enumerate}
\end{lemma'}

\begin{proof}
We prove part (1) by induction on the structure of the derivation of
$\termUV{\vec{x}; \delta; x_i}{M}$.  In the argument below, we let
${\cal D}$ be the derivation of 
$\lfprove{\Gamma, \Gamma_0, \Delta}{\oftype{M}{A}}$, and ${\cal D'}$ be
the derivation of $\lfprove{\Gamma, \Delta'}{\oftype{M'}{A'}}$.
\begin{packed_itemize}
\item
In the base case of \ruvinito, $M = x_i \vec{y}$
where $\vec{y}$ are distinct bound variables from $\delta$.
The derivation ${\cal D}$ must consist of $n$ \appobj\ rules and
a \varobj\ rule on $x_i$, whose type $B_i$ must be of the form
$\typedpis{z}{C}{D}$, with $A = D\subst{\vec{y}/\vec{z}}$.
Note that, because the variables $y_i$ are distinct bound variables
that are fresh with respect to $D$, this substitution can be inverted,
and we thus have $A\subst{\vec{z}/\vec{y}}=D$.
The other subderivations of the chain of \appobj\ applications
are instances of \varobj\ establishing
$\oftype{y_i}{C_i\subst{\vec{y}/\vec{z}}}$,
hence $(\oftype{y_i}{C'_i\subst{\vec{y}/\vec{z}}})\in\Delta'$
for $\encEq{C'_i}{C_i}$.

We next determine $t'_i$.
By $\eta$-equivalence we can assume that $t_i$ is of the form
$\lambda z_1 \ldots \lambda z_n. u$.
We have \[\encTerm{M'} = t_i \vec{y} = u \subst{\vec{y} / \vec{z}},\]
hence $u = \encTerm{M'}\subst{\vec{z} / \vec{y}} =
\encTerm{M' \subst{\vec{z} / \vec{y}}}$.
Let $u' = M' \subst{\vec{z} / \vec{y}}$ and
$t'_i = \typedlambdas{z}{C'}{u'}$.
We have 
\begin{eqnarray*}
\encTerm{t'_i} & = &
\lambda z_1 \ldots \lambda z_n.~ \encTerm{M'} \subst{\vec{z}
/ \vec{y}} \\
& = & \lambda z_1 \ldots \lambda z_n.~ u = t_i.
\end{eqnarray*}

We know that ${\cal D'}$ derives
$\lfprove{\Gamma, \Delta'}{\oftype{M'}{A'}}$.
From this we obtain a derivation of
\[\lfprove{\Gamma, 
\Delta'\subst{\vec{z}/\vec{y}}}{\oftype{u'}{A'\subst{\vec{z}/\vec{y}}}}\]
by renaming variables $\vec{y}$ into $\vec{z}$,
employing Proposition~\ref{prop:lf-renaming}.
The context $\Delta'\subst{\vec{z}/\vec{y}}$
contains assignments $(\oftype{z_i}{C'_i})$
and the other variables in its domain
do not occur in $u'$ nor $A' \subst{\vec{z}/\vec{y}}$
(since $\encEq{A'}{A}$, $A = D \subst{\vec{y}/\vec{z}}$ and $D$ is a subterm
  of $B_i$ which cannot contain any $y_i$).
We then have
\[\lfprove{\Gamma}{
          \oftype{(\typedlambdas{z}{C'}{u'})}{
                  (\typedpis{z}{C'}{
                               A'\subst{\vec{z}/\vec{y}}})}}\]
by weakening unused variables and using \absobj\ to introduce
the variables $\vec{z}$.
This is a typing derivation for $t'_i$; we must now show that
the associated type is actually the expected one:
\[B_i \subst{t'_1 / x_1 \ldots t'_{i-1} / x_{i-1}}\]
We have $\encTerm{A}\subst{t_1/x_1\ldots t_n/x_n} =
\encTerm{A\subst{t'_1/x_1\ldots t'_{i-1}/x_{i-1}}}$ and
$\encEq{A'}{A}$, from which we obtain, by injectivity of $\encTerm{\bullet}$,
that $A' = A\subst{t'_1/x'_1\ldots t'_{i-1}/x_{i-1}}$.
The same goes for $C'_i$ and $C_i$.
Since $B_i = \typedpis{z}{C}{A\subst{\vec{z}/\vec{y}}}$,
and the substitutions
$\subst{t'_1/x'_1\ldots t'_{i-1}/x_{i-1}}$ and $\subst{\vec{z}/\vec{y}}$
permute, we have:
\[ \typedpis{z}{C'}{A' \subst{\vec{z}/\vec{y}}} =
        B_i \subst{t'_1/x'_1\ldots t'_{i-1}/x_{i-1}} \]

\item
In the \ruvabso\ case, we have $M = \typedlambda{y}{A_1}{N}$ and ${\cal D}$
ends with the \absobj\ rule as follows:
\[
  \AxiomC{$\lfprove{\Gamma, \Gamma_0, \Delta}{\oftype{A_1}{\type}} \quad
           \lfprove{\Gamma, \Gamma_0, \Delta, \oftype{y}{A_1}}
                   {\oftype{N}{A_2}}$} 
  \UnaryInfC{
     $\lfprove{\Gamma, \Gamma_0, \Delta}
              {\oftype{(\typedlambda{y}{A_1}{N})}{(\typedpi{y}{A_1}{A_2})}}$} 
  \DisplayProof
\]
Then $\encEq{A'}{\typedpi{y}{A_1}{A_2}}$,
and hence $A'$ must be of the form $\typedpi{y}{A'_1}{A'_2}$ where
$\encEq{A'_i}{A_i}$.
Similarly, we obtain that $M'$ is of the form
$\typedlambda{y}{A'_1}{N'}$ with $\encEq{N'}{N}$.
Then, ${\cal D'}$ must contain a derivation of
\[\lfprove{\Gamma, \Delta', \oftype{y}{A'_1}}
         {\oftype{N'}{A'_2}},\]
and we conclude by the inductive hypothesis.

\item
In the \ruvappo\ case, we have $M = y \ N_1 \ldots N_m$,
$y \not\in \vec{x}$ and $\termUV{\vec{x}; \delta; x_i}{N_j}$.
Let $\typedpi{z_1}{C_1}{\ldots\typedpi{z_m}{C_m}{D}}$ be the type
of $y$ in $(\Gamma,\Delta)$.
The derivation ${\cal D}$ starts with a chain of \appobj\ applications,
followed by \varobj\ on $y$.
The premise corresponding to $N_j$ establishes that
\[ \lfprove{\Gamma, \Gamma_0, \Delta}{
        \oftype{N_j}{C_j \subst{N_1 / z_1, \ldots, N_{j-1} / z_{j-1}}}} \]
In $(\Gamma,\Delta')$, the variable $y$ is assigned the type
$\typedpis{z}{C'}{D'}$
with all $\encEq{C'_k}{C_k}$.
Moreover,
since $\encEq{M'}{(y \ N_1 \ldots N_m)}$ and since $y$ is not affected
by the instantiation of $\vec{x}$,
it must be that $M'$ is of the form $(y\app N'_1 \ldots N'_m)$
with all $\encEq{N'_j}{N_j}$.
The derivation ${\cal D'}$ must proceed in a similar fashion,
namely a chain of \appobj\ applications followed by \varobj\ on $y$.
Therefore we have a derivation of
\[\lfprove{\Gamma, \Delta'}
         {\oftype{N'_j}{C'_j \subst{N'_1 / z_1, \ldots, N'_{j-1} / z_{j-1}}}}\]
We can conclude by the inductive hypothesis because
\[\encEq{C'_j \subst{N'_1 / z_1 \ldots N'_{j-1} / z_{j-1}}}{
  C_j \subst{N_1 / z_1 \ldots N_{j-1} / z_{j-1}}}\] (which relies
on the disjointness of $\vec{x}$ and $\vec{z}$).
\end{packed_itemize}

The proof of (2) follows a similar pattern.
First, by a straightforward inspection of the first rules of
the derivation of
\[\lfprove{\Gamma}{\oftype{\typedpis{x}{B}{A}}{\type}}\]
we extract a derivation of $\lfprove{\Gamma, \Gamma_0}{\oftype{A}{\type}}$.
Then, since $A$ is a base type,
it must be (by rule \ruvappt) that
$x_i$ rigidly occurs in one of its arguments $M$.
Note that $A$ and $A'$ have the same structure on the path leading to $M$,
since no object is involved there.
Hence, a simultaneous inspection of the first rules of the derivations of
$\lfprove{\Gamma, \Gamma_0}{\oftype{A}{\type}}$
and
$\lfprove{\Gamma}{\oftype{A'}{\type}}$
yields derivations of
$\lfprove{\Gamma,\Gamma_0}{\oftype{M}{T}}$
and
$\lfprove{\Gamma}{\oftype{M'}{T'}}$ for
$\encEq{M'}{M}$ and $\encEq{T'}{T}$.
We can conclude using part (1).
\end{proof}

The definition of rigidity described above might seem restrictive.
In particular,
one might want to allow \[\termUV{\Gamma; \delta; x}{x \vec{N}}\] in \ruvinito.
However, with such a rule the rigidity lemma described above is no longer true.
For example, in a signature $\Gamma$ containing
$\oftype{num}{nat\ra\type}$ and
$\oftype{num_n}{\typedpi{n}{nat}{(num \app n)}}$,
the object $t = num_n$ provides a counter-example to 
Lemma~\ref{lemma:rigid-variables}, part~(1):
we have
$\lfprove{\Gamma, \oftype{x}{(nat\ra num\ z)}}{\oftype{(x \ z)}{(num \ z)}}$
and $\lfprove{\Gamma}{\oftype{(t \app z)}{(num \ z)}}$
but not $\lfprove{\Gamma}{\oftype{t}{nat \rightarrow num \app z}}$.
This example highlights a crucial aspect of our definition:
the applications allowed in \ruvinito\ should always
induce \emph{invertible} substitutions.
As in higher-order pattern
unification~\cite{miller91jlc,nipkow93lics}, we achieve this 
by restricting to applications involving a simple form of
$\beta$-reductions called {\em $\beta_0$-reductions} that are similar
to renaming.

We now use Lemma~\ref{lemma:rigid-variables} to prove the correctness
of the optimized translation.

\begin{theorem'}
  \label{theorem:extended-translation-equivalence}
  Let $\Gamma$ be an LF context, $A$ an LF type, both canonical,
  such that $\lfprove{}{\Gamma\ \ctx}$ and
  $\lfprove{\Gamma}{\oftype{A}{\type}}$ are derivable.
  Then when $M$ is an arbitrary \hhf\ term,
  $\iprove{\enc{\Gamma}}{\enc{A}(M)}$ has a derivation if
  and only if $\iprove{\encExtP{\Gamma}{}}{\encExtN{A}(M)}$ has a derivation.
\end{theorem'}

\begin{proof}
We establish the soundness direction by induction on the derivation
of the optimized translation, maintaining the assumptions about $\Gamma$
and $A$.

\medskip
\noindent If $A$ is of the form $\typedpi{x}{B}{A'}$
our derivation ends as follows:
\[
    \AxiomC{$\iprove{\encExtP{\Gamma, \oftype{x}{B}}{}}
                    {\encExtN{A'}(M \app x)}$}
    \RightLabel{\allGoal, \impGoal}
    \doubleLine
    \UnaryInfC{$\iprove{\encExtP{\Gamma}{}}
                       {\encExtN{\typedpi{x}{B}{A'}}(M)}$}
\DisplayProof
\]
First, $\lfprove{\Gamma}{\oftype{B}{\type}}$,
$\lfprove{}{(\Gamma, \oftype{x}{B}) \ \ctx}$
and 
$\lfprove{\Gamma, \oftype{x}{B}}{\oftype{A'}{\type}}$
must have derivations since $\Gamma$ and $A$ are well-formed.
We can thus apply the inductive hypothesis, obtaining that
\[\iprove{\enc{\Gamma, \oftype{x}{B}}}{\enc{A'}(M \app x)}\]
has a derivation.
By \allGoal\ and \impGoal{},
$\iprove{\enc{\Gamma}}{\enc{\typedpi{x}{B}{A'}}(M)}$
has one as well.

\medskip
\noindent If $A$ is a base type,
then our derivation starts with a backchaining on the encoding of some
$(\oftype{y}{\typedpis{x}{B}{A'}}) \in \Gamma$, \ie, on
\begin{tabbing}
\qquad\=\qquad\qquad\=\kill
\>$\forall x_1.~ (\encExtN{B_1}(x_1) \supset \ldots \supset$\\
\>\>$\forall x_n.~ (\encExtN{B_n}(x_n) \supset
     (u \ (y \ \vec{x}) \ \vec{\encTerm{N}}))).$
\end{tabbing}
In particular, this rule application has the form
\[
    \AxiomC{$\iprove{\encExtP{\Gamma}{}}{F_1}
             \quad \ldots \quad
             \iprove{\encExtP{\Gamma}{}}{F_n}$}
    \RightLabel{\bcGoal}
    \UnaryInfC{$\iprove{\encExtP{\Gamma}{}}
                       {(u (y \vec{x}) \vec{\encTerm{N}})
                         \subst{\vec{t / x}}}$}
\DisplayProof
\]
where $F_i$ is either
$(\encExtN{B_i }(x_i))\subst{t_1 / x_1, \ldots, t_{i} / x_{i}}$
or $\top$.
We perform an inner induction on $i\leq n$,
showing that for all $j\leq i$, $t_j = \encTerm{t'_j}$ for some
LF object $t'_j$, and that we have derivations of
\[\iprove{\enc{\Gamma}}
        {(\enc{B_j \subst{t'_1 / x_1, \ldots, t'_{j-1} / x_{j-1}}} \ t'_j)}\]
and
\[\lfprove{\Gamma}{\oftype{t'_j}{
    B_j \subst{t'_1 / x_1, \ldots, t'_{j-1} / x_{j-1}}}}.\]
\begin{packed_itemize}
\item
We first treat the case where $F_i=\top$, \ie,
there is a derivation of $\formulaUV{\vec{x}; x_i}{A'}$.
We assumed that
$\lfprove{\Gamma}{\oftype{A}{\type}}$,
and since $\Gamma$ is valid we also have a derivation of
$\lfprove{\Gamma}{\oftype{\typedpis{x}{B}{A'}}{\type}}$.
We can thus apply Lemma~\ref{lemma:rigid-variables}, to obtain $t'_i$ and
a derivation of
$\lfprove{\Gamma}
         {\oftype{t'_i}{B_i \subst{t'_1 / x_1, \ldots, t'_{i-1} / x_{i-1}}}}$,
and we conclude
by Theorem~\ref{theorem:simplified-translation-correctness}.
\item
When $F_i \neq \top$, we can see that within the
derivation of \[\lfprove{\Gamma}{\oftype{\typedpis{x}{B}{A'}}{\type}}\]
there is a derivation of
\[\lfprove{\Gamma, \oftype{x_1}{B_1}, \ldots, \oftype{x_{i-1}}{B_{i-1}}}
         {\oftype{B_i}{\type}}.\]
By substituting (Proposition~\ref{prop:lf-substitution}) the
derivations provided by the inner inductive hypothesis on this formula
we construct a derivation
of \[\lfprove{\Gamma}{
             \oftype{B_i \subst{t'_1/x_1,\ldots, t'_{i-1}/x_{i-1}}}{\type}}.\]
We can now apply the outer inductive hypothesis on $F_i$, to conclude that
$\iprove{\enc{\Gamma}}{
    (\enc{B_i \subst{t'_1 / x_1, \ldots, t'_{i-1} / x_{i-1}}}\ t_i)}$
has a derivation.
By Theorem~\ref{theorem:simplified-translation-correctness},
we finally obtain that $t_i$ is of the form $\encTerm{t'_i}$.
\end{packed_itemize}

We compose all derivations
\[ \iprove{\enc{\Gamma}}
        {\enc{B_i \subst{t'_1 / x_1, \ldots, t'_{i-1} / x_{i-1}}} \ t_i} \]
by \bcGoal\ on the encoding of 
$(\oftype{y}{\typedpis{x}{B}{A'}}) \in \Gamma$,
obtaining the expected derivation of
\[ \iprove{\enc{\Gamma}}{hastype \ (y \vec{t}) \
                         (u \ \vec{\encTerm{N}})\subst{\vec{t / x}}} \]

Completeness is proved by an induction
on the derivation of the simple translation.
This direction is rather straightforward as it consists only of dropping
information.
Details can be found in Appendix~\ref{appendix:proofs}.
\end{proof}

Therefore, by Theorems~\ref{theorem:simplified-translation-correctness} and
\ref{theorem:extended-translation-equivalence}, intuitionistic provability under the
optimized translation is equivalent to provability in LF, and the
following is a theorem.

\begin{theorem'}[Optimized translation correctness]
  \label{theorem:extended-translation-correctness}
  Let $\Gamma$ be an LF specification such that
  $\lfprove{}{\Gamma\ \ctx}$ has a
  derivation, $A$ an LF type such that
  $\lfprove{\Gamma}{\oftype{A}{\type}}$ has a derivation.
  Then, for any LF object $M$ such that
  $\lfprove{\Gamma}{\oftype{M}{A}}$ has a derivation,
  $\iprove{\encExtP{\Gamma}{}}{\encExtN{\oftype{M}{A}}}$ is derivable.
  Moreover, if
  $\iprove{\encExtP{\Gamma}{}}{\encExtN{A}(M)}$ for an arbitrary \hhf\ term $M$,
  then it must be that $M = \encTerm{M'}$ for some canonical LF object
  such that $\lfprove{\Gamma}{\oftype{M'}{A}}$ has a derivation.
\end{theorem'}

\section{Performance Comparisons}
\label{sec:results}

\begin{figure*}
\begin{minipage}{\textwidth}
\begin{center}
\begin{tabular}{lccccc}
Example           & Twelf & Simple   & Optimized & Typed Optimized & Indexing
\\[0pt] 
\hline
reverse(10)       & 1.0   & 0.40      & 0.14       &   0.07        & 0.08  \\
reverse(20)       & 1.0   & 0.57      & 0.19       &   0.12        & 0.11  \\
reverse(30)       & 1.0   & 0.63      & 0.20       &   0.14        & 0.11  \\
reverse(40)       & 1.0   & 0.41      & 0.13       &   0.10        & 0.07  \\
reverse(50)       & 1.0   & 0.46      & 0.15       &   0.10        & 0.08  \\
\hline
miniml(50)        & 1.0   &  0.74    &  0.25      &   0.18         & 0.08 \\
miniml(100)       & 1.0   &  1.25    &  0.44      &   0.30         & 0.17 \\
miniml(150)       & 1.0   &  1.75    &  0.56      &   0.41         & 0.25 \\
miniml(200)       & 1.0   &  2.89    &  0.83      &   0.62         & 0.41 \\
\hline
typed miniml(50)  & 1.0   & 2.27     & 1.07       &   0.57         & 0.48 \\
typed miniml(100) & 1.0   & 2.22     & 0.76       &   0.49         & 0.38 \\
typed miniml(150) & 1.0   & 3.49     & 1.44       &   0.67         & 0.55 \\
typed miniml(200) & 1.0   & 3.70     & 0.92       &   0.67         & 0.55 \\
\hline 
perm(10)          & 1.0   & \simerr  & 3.13      &   0.94          & 0.72 \\
perm(20)          & 1.0   & \simerr  & 1.75      &   0.78          & 0.44 \\
perm(30)          & 1.0   & \simerr  & 3.05      &   1.52          & 0.81 \\
perm(40)          & 1.0   & \simerr  & 3.95      &   2.15          & 1.14 \\
perm(50)          & 1.0   & \simerr  & 5.05      &   2.88          & 1.59 \\
\hline
num(64)           & 1.0   & 158.19   & 0.25      &   0.23          & 0.21 \\
num(128)          & 1.0   & $\infty$ & 0.10      &   0.10          & 0.07 \\
num(256)          & 1.0   & $\infty$ & 0.15      &   0.14          & 0.13 \\
num(512)          & 1.0   & $\infty$ & 0.003     &   0.003         & 0.003 \\
\end{tabular}
\end{center}
\end{minipage}
\caption{Performance comparison results}
\label{fig:comparisons}
\end{figure*}

We have claimed two properties for our translation: that it produces 
an \hhf\ program which corresponds closely to the original LF
specification, and that this program provides an
effective means for executing the specification. Evidence for the
first claim is provided by the translation of the $append$ 
specification presented in Figure~\ref{fig:optimized-append},
especially when one uses the easily applied simplification of a
formula of the form $\top \supset F$ to $F$. Notice also the
correspondence of the definition of the $append$ predicate to the one
that one might in, \eg\ Prolog, if one drops the first ``proof term''
argument of the predicate. To fully appreciate this benefit, it is
necessary to consider larger examples that space does not allow us to
do in this paper. However, such examples are available with the
implementation \cite{parinati.website}. We suggest that the reader
look especially at the example of the evaluator for Mini-ML with terms
that are not indexed by their type that is described below in the
collection of benchmarks: the translation
results in an \hhf\ program that is what one might
write in \hhf\ directly. 

\ignore{
The implementation of the translatation proceeds as follows.  The program
translates a given an LF specification and judgment into an \lprolog\ program.
This program is then compiled using the Teyjus implementation of \lprolog\
and run using the Teyjus virtual machine.
}

To test the second claim, we have carried out performance comparisons
between the Twelf implementation that interprets LF specifications
directly via a Standard ML program and an implementation obtained by
translating these specifications into \hhf\ programs and then
executing these using the Teyjus system. We
present results here over programs that have a few different characteristics:

\begin{packed_itemize}
\item
First, as we are interested in logic programming in LF, the traditional logic
program for naively reversing a list a $n$ times is included.
\item
The encoding of evaluators for various languages is a common usage of LF.
We have therefore used an encoding of Mini-ML along
with an encoding of addition as another sample program. This
benchmark, called \ital{miniml}, consists of adding $n$ to
$10$ using the encoding.
\item
The \ital{miniml} specification does not make essential use of
dependent types. The \ital{typed miniml}
benchmark, which consists of 
an evaluator for Mini-ML in which terms are indexed by their type,
uses dependent types to ensure that terms are well-formed.  The
Mini-ML program that was run is a typed version of the encoding
of addition. 
\item
An implementation of a meta-interpreter for intuitionistic
non-commutative linear logic (INCLL) has been proposed as a test
program \cite{pientka03cade}. The \ital{perm} benchmark
tests list permutation encoded in INCLL and run using the
meta-interpreter on lists of length $n$.
\item
The last benchmark, referred to as \ital{num}, involves rewriting arithmetic
expressions into an equivalent normal form. This example again makes
essential use of dependent types by associating with each equivalence
of two such terms a proof of their equivalence.  The benchmark tests
rewriting expressions of size $n$. 
\end{packed_itemize}

The third through fifth columns of Figure~\ref{fig:comparisons}
present data comparing  the simple translation, the translation with
redundant  typing judgments removed, and the fully optimized translation 
against the standard of Twelf with default optimizations on these
benchmarks.\footnote{This setting with Twelf
  leads to the best performance on these examples.} As
described in Figure~\ref{fig:optimized-translation}, the fully
optimized translation inserts the proof term as the first
argument of the predicate generated. Since this term is to be
determined by proof search, advantage cannot be taken of the
capability Teyjus possesses of 
indexing on the first argument. The last column presents data for the
case where we make the proof term the last argument instead. In the
data presented, \simerr\ indicates a heap overflow in the
Teyjus simulator, and $\infty$ means that the program ran more than
$1000$ times longer than Twelf.  

The most optimized translation
leads to better performance in most cases, often significantly so. On
the other hand, the simple translation yields a program that is
generally slower than Twelf. In particular, performance tends to
deteriorate with larger problems sizes, in keeping with the
difficulty  that we noted with this
translation. However, the simple translation is 
still  comparable to Twelf on the first three benchmarks.
On the \ital{perm} benchmark, Twelf does quite well and even
out-performs Teyjus with the optimized translation on problems of
large size. We have yet to pinpoint the reason for this---the program
is large and difficult to analyze in detail---but we suspect that the
linear head optimization that delays expensive unification computation
till after simpler checks have been made may have something to do with
this. The fact that term indexing causes significant improvement with
Teyjus gives credence to this observation.

For problems of very large size with all the benchmarks, the
performance of Twelf deteriorates quite dramatically; this is seen,
for example, in the case of \ital{num(n)} for a problem of size $512$.
This phenomenon is linked to the fact that Twelf consumes excessive
amounts of memory. The ultimate source of this problem is perhaps the
fact that Twelf is implemented in SML: it has been
argued that realizing a logic programming language in a functional
programming setting can lead to poor memory reclamation and eventually
to shortage of space  \cite{brisset94ilps}. 

\section{Conclusion and Future Work}
\label{sec:conclusion}

We have considered in this paper a translation of Twelf specifications
into logic programs in the \hhf\ language. An important part of our
ideas is the recognition of certain situations in which type
information is redundant in LF expressions and hence its checking can
be avoided. Our eventual translation produces a program that
corresponds closely to the original specification and we have argued that
it can be the basis for an effective animation of Twelf descriptions.

The specific work undertaken here can be extended in a few different
ways. As an extension to our notion of rigidity, we might observe that,
when applying a variable of type $\typedpis{x}{B}{A}$,
we could identify redundant type information, not only between a $B_i$
and $A$, but also between a $B_i$ and a different $B_j$.
It would also be interesting to relate our work to the ideas
of Reed~\cite{reed08tcs} who describes a notion of \emph{strictness}
similar to rigidity, used for the different purpose of identifying
sub-terms of LF objects that could be reconstructed if elided --
in contrast, we avoid redundant type checking but still generate
a complete LF object.
Such an understanding might lead both to an improvement of our
translation and to the ability to shorten LF terms that are needed in
applications such as that of proof-carrying-code \cite{necula97popl}.
From an implementation perspective, another possible optimization 
is to avoid constructing an LF object explicitly when the task has
been identified as that of only determining whether a type has an
inhabitant: experiments in this direction indicate in some cases a
ten-fold performance improvement over the optimized translation.
Techniques from the area of extracting programs from
proofs that pertain to isolating parts of a proof that do not
contribute to its overall computational content---\eg, see 
\cite{takayama91jsc}---are potentially useful to the application 
of such an optimization; these techniques might provide the basis for
noting components of a type whose inhabitants do not participate in
the term corresponding to the overall type.

We have focused here on realizing Twelf through a translation to
$\lambda$Prolog. A different approach, worthy of investigation, is
that of compiling Twelf specifications directly to bytecode for the
virtual machine underlying the Teyjus system. Such an approach would
make it possible to realize optimizations that have
been developed for the direct implementation of
Twelf~\cite{pientka06ijcar,pientka03cade}. Of special 
note here are optimizations like the linear heads treatment of unification
described by Pientka and Pfenning~\cite{pientka03cade} for minimizing
occurs checking, that could make a
difference in examples such as the \ital{perm} program considered in
the previous section: direct compilation would allow us to regain
opportunities for such improvements that might be lost by
translating first to $\lambda$Prolog and then relying on its 
implementation that is not specially optimized to treat
Twelf-specific programs. 

A more ambitious line of development concerns meta-reasoning over
specifications. 
Existing tools might be used to reason about LF programs
via the translation, the transparency of the translation becoming essential.
Anecdotal evidence
suggests that this transparency is not only enabling, it is also elucidating:
that the generated \hhf\ program is easier to reason about because
it highlights those types that could have logical importance, and elides
those that do not.

\section{Acknowledgements}

This work has been supported by the NSF grants CCR-0429572 and
CCF-0917140. Opinions, findings, and conclusions or recommendations
expressed in this papers are those of the authors and do not
necessarily reflect the views of the National Science Foundation.


\begin{thebibliography}{25}
\providecommand{\natexlab}[1]{#1}
\providecommand{\url}[1]{\texttt{#1}}
\expandafter\ifx\csname urlstyle\endcsname\relax
  \providecommand{\doi}[1]{doi: #1}\else
  \providecommand{\doi}{doi: \begingroup \urlstyle{rm}\Url}\fi

\bibitem[Baelde(2008)]{baelde08phd}
D.~Baelde.
\newblock \emph{A linear approach to the proof-theory of least and greatest
  fixed points}.
\newblock PhD thesis, Ecole Polytechnique, Dec. 2008.
\newblock URL \url{http://www.lix.polytechnique.fr/~dbaelde/thesis/}.

\bibitem[Baelde et~al.(2010)Baelde, Miller, and Snow]{baelde10ijcar}
D.~Baelde, D.~Miller, and Z.~Snow.
\newblock Focused inductive theorem proving.
\newblock In J.~Giesl and R.~Haehnle, editors, \emph{IJCAR}, Lecture Notes in
  Computer Science. Springer-Verlag, 2010.
\newblock (to appear).

\bibitem[Brisset and Ridoux(1994)]{brisset94ilps}
P.~Brisset and O.~Ridoux.
\newblock The architecture of an implementation of lambda-prolog: Prolog/mali.
\newblock In \emph{ILPS Workshop: Implementation Techniques for Logic
  Programming Languages}, 1994.

\bibitem[Church(1940)]{church40}
A.~Church.
\newblock A formulation of the simple theory of types.
\newblock \emph{J. of Symbolic Logic}, 5:\penalty0 56--68, 1940.

\bibitem[Felty(1989)]{felty89phd}
A.~Felty.
\newblock \emph{Specifying and Implementing Theorem Provers in a Higher-Order
  Logic Programming Language}.
\newblock PhD thesis, University of Pennsylvania, Aug. 1989.

\bibitem[Felty and Miller(1990)]{felty90cade}
A.~Felty and D.~Miller.
\newblock Encoding a dependent-type $\lambda$-calculus in a logic programming
  language.
\newblock In M.~Stickel, editor, \emph{Proceedings of the 1990 Conference on
  Automated Deduction}, volume 449 of \emph{LNAI}, pages 221--235. Springer,
  1990.

\bibitem[Gacek(2008)]{gacek08ijcar}
A.~Gacek.
\newblock The {A}bella interactive theorem prover (system description).
\newblock In A.~Armando, P.~Baumgartner, and G.~Dowek, editors, \emph{Fourth
  International Joint Conference on Automated Reasoning}, volume 5195 of
  \emph{LNCS}, pages 154--161. Springer, 2008.
\newblock URL \url{http://arxiv.org/abs/0803.2305}.

\bibitem[Gacek(2009)]{gacek09phd}
A.~Gacek.
\newblock \emph{A Framework for Specifying, Prototyping, and Reasoning about
  Computational Systems}.
\newblock PhD thesis, University of Minnesota, 2009.

\bibitem[Gacek et~al.(2008{\natexlab{a}})Gacek, Holte, Nadathur, Qi, and
  Snow]{teyjus.website}
A.~Gacek, S.~Holte, G.~Nadathur, X.~Qi, and Z.~Snow.
\newblock The {T}eyjus system -- version 2, Mar. 2008{\natexlab{a}}.
\newblock Available from \url{http://teyjus.cs.umn.edu/}.

\bibitem[Gacek et~al.(2008{\natexlab{b}})Gacek, Miller, and
  Nadathur]{gacek08lics}
A.~Gacek, D.~Miller, and G.~Nadathur.
\newblock Combining generic judgments with recursive definitions.
\newblock In F.~Pfenning, editor, \emph{23th Symp.\ on Logic in Computer
  Science}, pages 33--44. IEEE Computer Society Press, 2008{\natexlab{b}}.

\bibitem[Harper et~al.(1993)Harper, Honsell, and Plotkin]{harper93jacm}
R.~Harper, F.~Honsell, and G.~Plotkin.
\newblock A framework for defining logics.
\newblock \emph{Journal of the ACM}, 40\penalty0 (1):\penalty0 143--184, 1993.

\bibitem[Howard(1980)]{howard80}
W.~A. Howard.
\newblock The formulae-as-type notion of construction.
\newblock In J.~P. Seldin and R.~Hindley, editors, \emph{To H. B. Curry: Essays
  in Combinatory Logic, Lambda Calculus, and Formalism}, pages 479--490.
  Academic Press, New York, 1980.

\bibitem[Miller(1991)]{miller91jlc}
D.~Miller.
\newblock A logic programming language with lambda-abstraction, function
  variables, and simple unification.
\newblock \emph{J. of Logic and Computation}, 1\penalty0 (4):\penalty0
  497--536, 1991.

\bibitem[Miller and Tiu(2005)]{miller05tocl}
D.~Miller and A.~Tiu.
\newblock A proof theory for generic judgments.
\newblock \emph{ACM Trans.\ on Computational Logic}, 6\penalty0 (4):\penalty0
  749--783, Oct. 2005.

\bibitem[Miller et~al.(1991)Miller, Nadathur, Pfenning, and
  Scedrov]{miller91apal}
D.~Miller, G.~Nadathur, F.~Pfenning, and A.~Scedrov.
\newblock Uniform proofs as a foundation for logic programming.
\newblock \emph{Annals of Pure and Applied Logic}, 51:\penalty0 125--157, 1991.

\bibitem[Nadathur and Miller(1988)]{nadathur88iclp}
G.~Nadathur and D.~Miller.
\newblock An {Overview} of {$\lambda$Prolog}.
\newblock In \emph{{Fifth International Logic Programming Conference}}, pages
  810--827, Seattle, Aug. 1988. MIT Press.

\bibitem[Necula(1997)]{necula97popl}
G.~C. Necula.
\newblock Proof-carrying code.
\newblock In \emph{Conference Record of the 24th Symposium on Principles of
  Programming Languages 97}, pages 106--119, Paris, France, 1997. ACM Press.

\bibitem[Nipkow(1993)]{nipkow93lics}
T.~Nipkow.
\newblock Functional unification of higher-order patterns.
\newblock In M.~Vardi, editor, \emph{Proc.\ 8th {IEEE} Symposium on Logic in
  Computer Science ({LICS} 1993)}, pages 64--74. IEEE, June 1993.

\bibitem[Pfenning and Sch{\"u}rmann(1999)]{pfenning99cade}
F.~Pfenning and C.~Sch{\"u}rmann.
\newblock System description: Twelf --- {A} meta-logical framework for
  deductive systems.
\newblock In H.~Ganzinger, editor, \emph{16th Conference on Automated Deduction
  (CADE)}, number 1632 in LNAI, pages 202--206, Trento, 1999. Springer.

\bibitem[Pientka(2006)]{pientka06ijcar}
B.~Pientka.
\newblock Eliminating redundancy in higher-order unification: A lightweight
  approach.
\newblock In U.~Furbach and N.~Shankar, editors, \emph{IJCAR}, volume 4130 of
  \emph{Lecture Notes in Computer Science}, pages 362--376. Springer, 2006.
\newblock ISBN 3-540-37187-7.

\bibitem[Pientka and Pfenning(2003)]{pientka03cade}
B.~Pientka and F.~Pfenning.
\newblock Optimizing higher-order pattern unification.
\newblock In \emph{19th International Conference on Automated Deduction}, pages
  473--487. Springer-Verlag, 2003.

\bibitem[Reed(2008)]{reed08tcs}
J.~Reed.
\newblock Redundancy elimination for {LF}.
\newblock \emph{Electron. Notes Theor. Comput. Sci.}, 199:\penalty0 89--106,
  2008.
\newblock ISSN 1571-0661.
\newblock \doi{http://dx.doi.org/10.1016/j.entcs.2007.11.014}.

\bibitem[Snow(2010{\natexlab{a}})]{parinati.website}
Z.~Snow.
\newblock {P}arinati.
\newblock \url{http://www.cs.umn.edu/~snow/parinati}, 2010{\natexlab{a}}.

\bibitem[Snow(2010{\natexlab{b}})]{snow10masters}
Z.~Snow.
\newblock Realizing the dependently typed $\lambda$-calculus.
\newblock Master's thesis, University of Minnesota, 2010{\natexlab{b}}.

\bibitem[Takayama(1991)]{takayama91jsc}
Y.~Takayama.
\newblock Extraction of redundancy-free programs from constructive natural
  deduction proofs.
\newblock \emph{Journal of Symbolic Computation}, 12\penalty0 (1):\penalty0
  29--69, 1991.

\end{thebibliography}

\appendix

\section{Proofs of Theorems} \label{appendix:proofs}

\subsection{Correctness of the simplified encoding\\
  (Theorem~\ref{theorem:simplified-translation-correctness})}

\subsubsection{Completeness}

We use induction on the derivation
of $\lfprove{\Gamma}{\oftype{M}{A}}$ to build one for
$\iprove{\enc{\Gamma}}{\enc{\oftype{M}{A}}}$.
We proceed by case analysis on the canonical type $A$.

\smallskip
\noindent If $A$ is of the form $\typedpi{x}{B}{A'}$ then $M$ must be
of the form $\typedlambda{x}{B}{M'}$ and the LF derivation
must end with an \absobj\ rule, \ie, a rule of the form
\[
   \AxiomC{$\lfprove{\Gamma}{\oftype{A'}{\type}} \quad
     \lfprove{\Gamma,\oftype{x}{B}}{\oftype{M}'{A'}}$}
   \RightLabel{\absobj}
   \UnaryInfC{$\lfprove{\Gamma}{
      \oftype{(\typedlambda{x}{B}{M'})}{(\typedpi{x}{B}{A'})}}$}
   \DisplayProof
\]
The induction hypothesis gives us a derivation for
\begin{tabbing}
\qquad\=\kill
\>$\iprove{\enc{\Gamma, \oftype{x}{B}}}{\enc{\oftype{M'}{A'}}}$.
\end{tabbing}
By applying the rules \allGoal\ and \impGoal\ to this, we get a
derivation for 
$\iprove{\enc{\Gamma}}{
    \forall x.~ \enc{\oftype{x}{B}} \supset \enc{\oftype{M'}{A'}}}$.
The righthand side of this sequent is the expected goal:
\begin{tabbing}
\qquad\=\qquad\qquad\=\kill
\>$\enc{\oftype{(\typedlambda{x}{B}{M'})}{(\typedpi{x}{B}{A'})}} =$\\
\>\>$\forall x.~ \enc{\oftype{x}{B}} \supset
(\enc{A'} \ (\encTerm{\typedlambda{x}{B}{M'}} \ x))$,
\end{tabbing}
and $\encTerm{M'} = (\encTerm{\typedlambda{x}{B}{M'}}\app x)$ by virtue of
$\eta$-conversion. 

\medskip
\noindent If $A$ is a base type then $M$ must be of the form
$x\app N_1 \app \ldots \app N_n$ and the canonical LF derivation must
end with a chain of \appobj\ rules following a \varobj\ rule that
reveals that 
\begin{tabbing}
\qquad\=\kill
\>$\oftype{x}{\typedpi{y_1}{B_1}{\ldots\typedpi{y_n}{B_n}{A'}}}\in\Gamma$.
\end{tabbing}
Moreover, $A$ must be $A'\subst{{N_1/y_1,\ldots,N_n/y_n}}$ and, from looking at the right
upper premise of the \appobj\ rules, there must be shorter derivations
of 
\begin{tabbing}
\qquad\=\kill
\>$\lfprove{\Gamma}{\oftype{N_i}{B_i\subst{N_1/x_1,\ldots,N_{i-1}/x_{i-1}}}}$
\end{tabbing}
for $1 \leq i \leq n$. By the induction hypothesis we obtain
derivations ${\cal D}_i$ of $\iprove{\enc{\Gamma}}{ 
   \enc{\oftype{N_i}{B_i\subst{N_1/x_1, \ldots, N_{i-1}/x_{i-1}}}}}$.
Further, $\enc{\Gamma}$ must contain 
\begin{tabbing}
\qquad\=\qquad\=\kill 
\>$\forall y_1.~ (\enc{B_1} \app y_1) \supset \ldots \supset$\\
\>\>$\forall y_n.~ (\enc{B_n} \app y_n) \supset
     hastype \app (x \app y_1 \app \ldots\app y_n) \app \encTerm{A'}$,
\end{tabbing}
\ie, the encoding of
     $\oftype{x}{\typedpi{y_1}{B_1}{\ldots\typedpi{y_n}{B_n}{A'}}}$.
By applying \bcGoal\ on that clause,
choosing $\encTerm{N_i}$ for $y_i$
and using the derivations ${\cal D}_i$,
we obtain a derivation of
\begin{tabbing}
\qquad\=$\iprove{\enc{\Gamma}}{hastype \app}$\=\kill
\>$\iprove{\enc{\Gamma}}{hastype \app
(x \app \encTerm{N_1}\app\ldots\app \encTerm{N_n})$\\
\>\>$(\encTerm{A'}\subst{\encTerm{N_1}/y_1,\ldots
                           \encTerm{N_n}/y_n})}$.
\end{tabbing}
The right side of this sequent is precisely
\begin{tabbing}
\qquad\=\kill
\>$\enc{\oftype{(x \app N_1 \app\ldots\app
N_n)}{A'\subst{N_1/y_1,\ldots,N_n/y_n}}}$.  
\end{tabbing}

\subsubsection{Soundness}

We prove the soundness direction by induction on the derivation of
$\iprove{\enc{\Gamma}}{(\enc{A}\app M)}$: assuming that
$\lfprove{\Gamma}{\oftype{A}{\type}}$ has a derivation,
we establish that $M = \encTerm{M'}$ for some canonical object $M'$
and we build a derivation of $\lfprove{\Gamma}{\oftype{M'}{A}}$.
A case analysis on the structure of the canonical type $A$
will guide us.

\smallskip
\noindent If $A$ is of the form $\typedpi{x}{B}{A'}$ then the structure of
$\enc{A}$ forces the \hhf\ derivation to conclude as follows:
\[
  \AxiomC{$
    \iprove{\enc{\Gamma,\oftype{x}{B}}}{(\enc{A'}\app (M\app x))}
  $}
  \RightLabel{\allGoal, \impGoal}
  \doubleLine
  \UnaryInfC{$\iprove{\enc{\Gamma}}{
     \forall x.~ (\enc{B}\app x) \supset (\enc{A'}\app (M\app x))}$}
  \DisplayProof
\]
Since $A$ is a valid $\type$ under $\Gamma$, $B$ must also be,
and $A'$ must be valid under $(\Gamma,\oftype{x}{B})$.
We can thus apply the inductive hypothesis,
and we obtain that $M\app x = \encTerm{M'}$ 
and that $\lfprove{\Gamma,\oftype{x}{B}}{\oftype{M'}{A'}}$ is
derivable for some canonical object $M'$.
Since $x$ does not occur free in $M$, we conclude that
\begin{tabbing}
\qquad\=\kill
\>$M = (\lambda x. \encTerm{M'}) = \encTerm{\typedlambda{x}{B}{M'}}$,
\end{tabbing}
and we derive
$\lfprove{\Gamma}{\oftype{(\typedlambda{x}{B}{M'})}{(\typedpi{x}{B}{A'})}}$
using the \absobj\ rule and our derivation of
$\lfprove{\Gamma}{\oftype{B}{\type}}$.

\medskip
\noindent
Otherwise, $A$ is a base type, and the derivation we are considering
is that of $\iprove{\enc{\Gamma}}{hastype\app
M\app \encTerm{A}}$. This derivation must end in a \bcGoal\ rule that
uses some clause in $\enc{\Gamma}$ of the form
\begin{tabbing}
\qquad\=\qquad\=\kill
\>$\forall y_1.~ (\enc{B_1}\app y_1) \supset \ldots \supset$\\
\>\>$\forall y_n.~ (\enc{B_n} \app y_n) \supset
     hastype \app (x \app y_1 \app\ldots\app y_n) \app \encTerm{A'}$;
\end{tabbing}
note that the variables $y_1,\ldots,y_{i-1}$ can appear in $\enc{B_i}$
here. 
Thus, for some \hhf\ terms $N_1,\ldots,N_n$, 
\begin{tabbing}
\qquad\=\kill
\>$\encTerm{A} = \encTerm{A'}\subst{N_1/y_1,\ldots,N_n/y_n}$,
\end{tabbing}
$M = (x\app N_1\app \ldots\app N_n)$, and, for each $i$ such that
$1 \leq i \leq n$, there is a shorter derivation of
\begin{tabbing}
\qquad\=\kill
\>$\iprove{\enc{\Gamma}}{(\enc{B_i} \
y_i)\subst{N_1/y_1,\ldots,N_i/y_i}}$,
\end{tabbing}
\ie, of $\iprove{\enc{\Gamma}}{
           (\enc{B_i}\subst{N_1/y_1,\ldots,N_{i-1}/y_{i-1}}\app N_i})$.
Further, we know that
$\oftype{x}{\typedpi{y_1}{B_1}{\ldots\typedpi{y_n}{B_n}A'}}\in\Gamma$ for 
some $x$. We now claim that, for $1 \leq i\leq n$,
$N_i = \encTerm{N'_i}$ for some canonical LF object $N'_i$ and
that $\lfprove{\Gamma}{\oftype{N'_i}{B_i \subst{N'_1/y_1\ldots
N'_{i-1}/y_{i-1}}}}$ has a derivation. If this claim is true, then,
we can use the \varobj\ rule to derive
$\lfprove{\Gamma}{\oftype{x}{\typedpi{y_1}{B_1}{\ldots\typedpi{y_n}{B_n}A'}}}$
and follow this by a sequence of \appobj\ rule applications to prove
\[\lfprove{\Gamma}{\oftype{(x \app N'_1\app \ldots\app N'_n)}{
                          A'\subst{N'_1/y_1 \ldots N'_n/y_n}}}.\]
Now, evidently $M = \encTerm{x\app N'_1\app\ldots\app N'_n}$ and,
since substitution permutes with encoding, $A = A'\subst{N'_1/y_1,\ldots,N'_n/y_n}$. Thus, the desired result would
be proven. 

It only remains to establish the claim. We actually strengthen
it to include also the assertion that, for $1 \leq i \leq
n$, 
\[\lfprove{\Gamma}{\oftype{B_i\subst{N'_1/y_1 \ldots
N'_{i-1}/y_{i-1}}}{\type}}\] has a derivation. To prove it, we use an 
inner induction on 
$i$. Since $\Gamma$ is a well-formed context, and
$\oftype{x}{\typedpi{y_1}{B_1}{\ldots\typedpi{y_n}{B_n}A'} \in \Gamma$,
there must be a derivation of 
\[\lfprove{\Gamma,\oftype{x_1}{B_1},\ldots,\oftype{x_{i-1}}{B_{i-1}}}
            {\oftype{B_i}{\type}}\] 
for $1 \leq i \leq n$. Using Proposition~\ref{prop:lf-substitution}
and the induction hypothesis we see that there must be a
derivation of 
\[\lfprove{\Gamma}{\oftype{B_i\subst{N'_1/y_1 \ldots 
      N'_{i-1}/y_{i-1}}}{\type}}.\]
Noting that \[\enc{B_i}\subst{N_1/y_1,\ldots,N_{i-1}/y_{i-1}}}
      = \enc{B_i\subst{N_1/y_1,\ldots,N_{i-1}/y_{i-1}}},\] 
the outer induction hypothesis and the shorter derivation
      of 
\[\iprove{\enc{\Gamma}}{
           (\enc{B_i}\subst{N_1/y_1,\ldots,N_{i-1}/y_{i-1}}\app N_i})\]
          allows us to conclude that $N_i = \encTerm{N'_i}$ for some
          canonical LF term $N'_i$ and that there is a derivation
           of \[\lfprove{\Gamma}{\oftype{N'_i}{B_i \subst{N'_1/y_1\ldots
N'_{i-1}/y_{i-1}}}},\] thus verifying the claim. 

\subsection{Completeness of the optimized encoding
  (Theorem~\ref{theorem:extended-translation-equivalence})}
                                   
If $\iprove{\enc{\Gamma}}{\enc{A} M}$ has a derivation, then
$\iprove{\encExtP{\Gamma}{}}{\encExtN{A} M}$ has a derivation as well.
Note that for this direction of the proof we are simply dropping
information (subderivations) and so we do not
rely on $\Gamma$ being a valid specification or
$A$ being a valid type.
We proceed by induction on the structure of the derivation of
$\iprove{\enc{\Gamma}}{\enc{A} M}$, followed by case analysis on $A$.

\smallskip
\noindent 
If $A$ is of the form $\typedpi{x}{B}{A'}$ our derivation ends as follows:
\[
    \AxiomC{$\iprove{\enc{\Gamma, \oftype{x}{B}}}{\enc{A'} \ (M \app x)}$}
    \RightLabel{\allGoal, \impGoal}
    \doubleLine
    \UnaryInfC{$\iprove{\enc{\Gamma}}
                       {\enc{\typedpi{x}{B}{A'}} \ M}$}
\DisplayProof
\]
By the inductive hypothesis
$\iprove{\encExtP{\Gamma, \oftype{x}{B}}{}}{\encExtN{A'} \ (M \app x)}$ has a
derivation, and by applying \allGoal\ and \impGoal\ to this derivation we can
construct a derivation of
\[ \iprove{\encExtP{\Gamma}{}}
        {\encExtN{\typedpi{x}{B}{A'}} \ M} \]

\smallskip
\noindent
Otherwise, $A$ is a base type and our derivation proceeds by backchaining
on some $(\oftype{y}{\typedpis{x}{B}{A'}})\in\Gamma$,
with $\encTerm{A}=\encTerm{A'}\subst{t_1/x_1\ldots t_n/x_n}$:
\[
    \AxiomC{$\iprove{\enc{\Gamma}}{F_1}
             \quad \ldots \quad
             \iprove{\enc{\Gamma}}{F_n}$}
    \RightLabel{\bcGoal}
    \UnaryInfC{$\iprove{\enc{\Gamma}}
                       {\enc{A} \ (y \vec{t})}$}
\DisplayProof
\]
Here, $F_i = (\enc{B_i} \ x_i)\subst{t_1/x_1\ldots t_n/x_n}$.
As in the completeness proof of the simplified encoding,
we obtain by an inner induction that each $t_i$ is of the form
$\encTerm{t'_i}$ and thus that
\[F_i = \enc{B_i\subst{t'_1/x_1\ldots t'_n/x_n}}(t_i).\]
We shall build the derivation of
$\iprove{\encExtP{\Gamma}{}}{\encExtN{A}(y \vec{t})}$
by using \bcGoal\ on the optimized encoding of
$(\oftype{y}{\typedpis{x}{B}{A'}})\in\Gamma$,
by choosing $\vec{t}$ for $\vec{x}$.
The resulting premises are either
\[ \iprove{\encExtP{\Gamma}{}}{
   \encExtN{B_i\subst{t'_1/x_1\ldots t'_n/x_n}} \ t_i} \]
when $x_i$ does not occur rigidly in $A'$,
and this case is provided for by the inductive hypothesis,
or $\top$ otherwise, which we derive using \topGoal.

\end{document}